\documentclass[12pt,a4paper]{article}
\usepackage[latin1]{inputenc}
\usepackage{amsmath}
\usepackage{amsfonts}
\usepackage{amssymb}
\usepackage{graphicx}
\usepackage{tikz}
\DeclareRobustCommand\full  {\tikz[baseline=-0.6ex]\draw[red,thick] (0,0)--(0.5,0);}
\DeclareRobustCommand\dotted{\tikz[baseline=-0.6ex]\draw[orange,thick,dotted] (0,0)--(0.54,0);}
\DeclareRobustCommand\dashed{\tikz[baseline=-0.6ex]\draw[blue,thick,dashed] (0,0)--(0.54,0);}
\DeclareRobustCommand\chain {\tikz[baseline=-0.6ex]\draw[black,thick,dash dot dot] (0,0)--(0.5,0);}
\usepackage{caption}
\usepackage{subcaption}
\usepackage{sidecap}
\usepackage{color}
\usepackage{chngcntr}
\usepackage{cite}
\usepackage{authblk}
\counterwithout{figure}{section}
\def\vecsign{\mathchar"017E}
\def\dvecsign{\smash{\stackon[-1.95pt]{\vecsign}{\rotatebox{180}{$\vecsign$}}}}
\def\dvec#1{\def\useanchorwidth{T}\stackon[-4.2pt]{#1}{\,\dvecsign}}
\usepackage{stackengine}
\stackMath
\title{Barut-Girardello coherent states for anisotropic 2D-Dirac materials}
\date{ }
\author[1]{E D\'iaz-Bautista\footnote{ediaz@fis.cinvestav.mx}}
\author[2]{Y Concha-S\'anchez\footnote{yconcha@umich.mx}}
\author[3]{A Raya\footnote{raya@ifm.umich.mx}}
\affil[1]{\small Physics Department, Cinvestav, P.O. Box. 14-740, 07000 Mexico City, Mexico}
\affil[2]{\small Facultad de Ingenier\'{\i}a Civil, Universidad Michoacana de San Nicol\'as de Hidalgo, Edificio C, Ciudad Universitaria. Francisco J. M\'ujica s/n. Col. Fel\'{\i}citas del R\'{\i}o. 58030, Morelia, Michoac\'an, M\'exico}
\affil[3]{\small Instituto de F\'{\i}sica y Matem\'aticas, Universidad Michoacana de San Nicol\'as de Hidalgo, Edificio C-3, Ciudad Universitaria. Francisco J. M\'ujica s/n. Col. Fel\'{\i}citas del R\'{\i}o. 58040 Morelia, Michoac\'an, M\'exico}
\begin{document}
	\maketitle

\begin{abstract}
We construct the Barut-Girardello coherent states for charge carriers in anisotropic 2D-Dirac materials immersed in a constant homogeneous magnetic field which is orthogonal to the sample surface. For that purpose, we solve the anisotropic Dirac equation and identify the appropriate arising and lowering operators. Working in a Landau-like gauge, we explicitly construct nonlinear coherent states as eigenstates of a generalized annihilation operator with complex eigenvalues which depends on an arbitrary function $f$ of the number operator. In order to describe the anisotropy effects on these states, we obtain the Heisenberg uncertainty relation, the probability density, mean energy value and occupation number distribution for three different functions $f$. 
For the case in which the anisotropy is caused by uniaxial strain, we obtain that when a stress is applied along the $x$-axis of the material surface, the probability density for the nonlinear coherent states is smaller compared to when the material is stressed along the orthogonal axis.
\end{abstract}

\section{Introduction}\label{intro}
The physical system of a charged particle interacting with a uniform magnetic field  has been considered in several works due to its important technological implications. Fock solved the non-relativistic quantum mechanical problem for the first time by defining the magnetic field in the so-called symmetric gauge~\cite{f28}, but Landau addressed the same physical situation by choosing a gauge --nowadays known as Landau gauge-- that reduces the initial Schr\"odinger equation to the one-dimensional quantum harmonic oscillator problem~\cite{l30}. Although trivial at first glance, this fact allows to connect with a well-known system that can be solved algebraically by defining a set of first order differential operators $a$ and $a^\dagger$, that together with the identity operator are generators of the Heisenberg-Weyl (HW) algebra. For the quantum harmonic oscillator, the coherent states describe  such a system in semi-classical situations. In fact, Schr\"{o}dinger~\cite{s26} proposed the coherent states (CS) as the most classical states describing the motion of a particle in a quadratic potential, and every since then, they have become a canonical subject in quantum mechanics literature. Among many other advantages CS have been used to test, both experimentally and theoretically, features of interferometry in many branches of physics, ranging from optics, atomic, nuclear, condensed matter and particle physics (see, for example, Ref.~\cite{ks85} and references therein). Moreover, the construction of coherent states has been generalized to other systems through different definitions, e.g., as eigenstates of the annihilation operator of the system (Barut-Girardello CS)~\cite{bg71}, or as states obtained by acting the displacement operator on the fundamental state (Gilmore-Perelomov CS)~\cite{p72,p86,g72,g74}.


The algebra associated to the arising and lowering operators $a^\dagger$, $a$ of the harmonic oscillator can be generalized to an $f$-deformed algebra, which is obtained by replacing them by deformed creation and annihilation  operators defined as~\cite{mmzs96}
\begin{equation}
\mathcal{A}=af(N)=f(N+1)a, \quad \mathcal{A}^\dagger=f(N)a^\dagger=a^\dagger f(N+1),
\end{equation}
where $f$ is a well-behaved real function of the standard number operator $N=a^\dagger a$, with the corresponding commutators
\begin{equation}\label{2}
[N,\mathcal{A}]=-\mathcal{A}, \quad [N,\mathcal{A}^\dagger]=\mathcal{A}^\dagger, \quad [\mathcal{A},\mathcal{A}^+]=(N+1)f^2(N+1)-Nf^2(N).
\end{equation}
Thus, nonlinear coherent states (NLCS) have been introduced as eigenstates of the deformed annihilation operator $\mathcal{A}\vert\alpha\rangle_f=\alpha\vert\alpha\rangle_f$~\cite{mmzs96,mmsz97}. In general, such states exhibit nonclassical properties, e.g., squeezing and antibunching~\cite{mv96}. They are also connected with oscillators whose frecuency depends on the energy~\cite{mmzs96,mmsz97,ss00}, some of them can be obtained physically as stationary states of the center-of-mass motion of a trapped ion~\cite{mv96} or to model the vibrations of polyatomic molecules~\cite{bd92,bd92i}. and more. Hence, it can be concluded that the construction of coherent states for a quantum mechanical system is a desirable thing to do.

On the other hand, the so-called 2D-Dirac materials (2D-DM), such as graphene \cite{nmf06,k07,njzm07,gn07}, topological insulators \cite{hk10,qz11} and organic conductors \cite{kks09,kntsk14}, are characterized because, at low-energy ({\em i.e.,} in the continuum limit), the behavior of its charge carriers is quite similar to that of ultra-relativistic fermions, because its dispersion relation is linear.~As a consequence, these quasiparticles are described by a Dirac-like equation, instead of the ordinary Schr\"{o}dinger equation with a typical parabolic dispersion relation.~Several phenomena related to the pseudo-relativistic behavior of these quasiparticles have been studied extensively, for example, in graphene --the most-known 2D material-- in response to applied external magnetic fields due to its outstanding properties for technological applications and fundamental physics development.

Recently, an increasing interest to exploit strain for controlling other physical properties of the 2D-DM, {e.g.}, their stiffness, strength and optical conductivity has arisen due to their mechanical properties \cite{ow17}. For example, among the new research subjects worth to be mentioned, straintronics~\cite{pcp09} studies the mechanical deformations of graphene layers to modify its electric properties~\cite{gbot17}. Actually, some experimental results regarding the response of graphene under tensile and compressive strain have been discussed previously~\cite{tppj09}. Theoretically, although these mechanical deformations displace and deform the Dirac cones to an elliptic cross-section and induce a tensor character to the Fermi velocity, the equations of motion are still tractable~\cite{gbot17}. However, despite the simplicity that the assumption of certain types of deformations in graphene~\cite{og13,bccc15,ow17-1} could offer, our goal here is to generalize the results in~\cite{ed17} towards the anisotropic Dirac fermion systems by constructing the corresponding NLCS in order to give a semi-classical description of the phenomena related with the combined effects of both magnetic fields and anisotropy, and that later allow to analyze other interesting physical properties of these materials~\cite{ks85,fk70,acg72,wh73,zfg90,aag00}.
For that purpose, we have organized this article as follows. In sect.~\ref{dirac} the anisotropic 2D-Dirac equation is solved in a Landau-like gauge. The corresponding energy spectrum and eigenstates are obtained as functions of a parameter $\zeta$ that characterizes the {anisotropy}. In sect.~\ref{annihilation} a generalized annihilation operator associated to the system is presented and the NLCS are introduced as eigenstates of such a matrix operator. These quantum states are characterized through their probability density, the Heisenberg uncertainty relation and the mean energy value. In sect.~\ref{conclu} we discuss our achievements and, as an example, we present our conclusions for the strained graphene case.

\section{Anisotropic 2D-Dirac Hamiltonian}\label{dirac}
Let us recall that, departing from the pristine case, the 2D-Dirac Hamiltonian
	\begin{equation}
		H=v_F \vec{\sigma}\cdot \vec p,
	\end{equation}
	where $\vec{\sigma}=(\sigma_x,\sigma_y)$ denotes the Pauli matrices, may be modified either because the material is inherently anisotropic or has been altered through mechanical deformations, giving as a result in both cases that the Fermi-velocity $v_F$ is no longer isotropic. This fact is accounted for by modifying the anisotropic Hamiltonian as
	\begin{equation}
		H=v_F \vec{\sigma}\cdot {\vec p}',
	\end{equation}
	where ${\vec p}'$ is the momentum measured from the new Dirac points and is related with $\vec p$ as
	\begin{equation}
		{\vec p}'=R(\theta)S(\epsilon)R(-\theta)\vec p,
	\end{equation}
where the matrix $R(\theta)$ represents a rotation along the anisotropy direction and $S(\epsilon)$ describes the deformation of the Dirac cones due to it. For the case of a strain of strength $\epsilon$ applied uniaxially see Ref.~\cite{pap11}, for instance. Thus, a number of physical observables for the pristine and anisotropic cases are linearly related through transformations involveng these matrices. Such is the case, for instance, of linear response correlation functions, which are related as~\cite{pap11}
\begin{equation}
	\Pi({\vec p}')=({\rm det}~S(\epsilon))^{-1} \Pi({\vec p}).
\end{equation}

\begin{figure}[h!]
	\centering
	\begin{minipage}[b]{0.45\textwidth}
		\includegraphics[width=\textwidth]{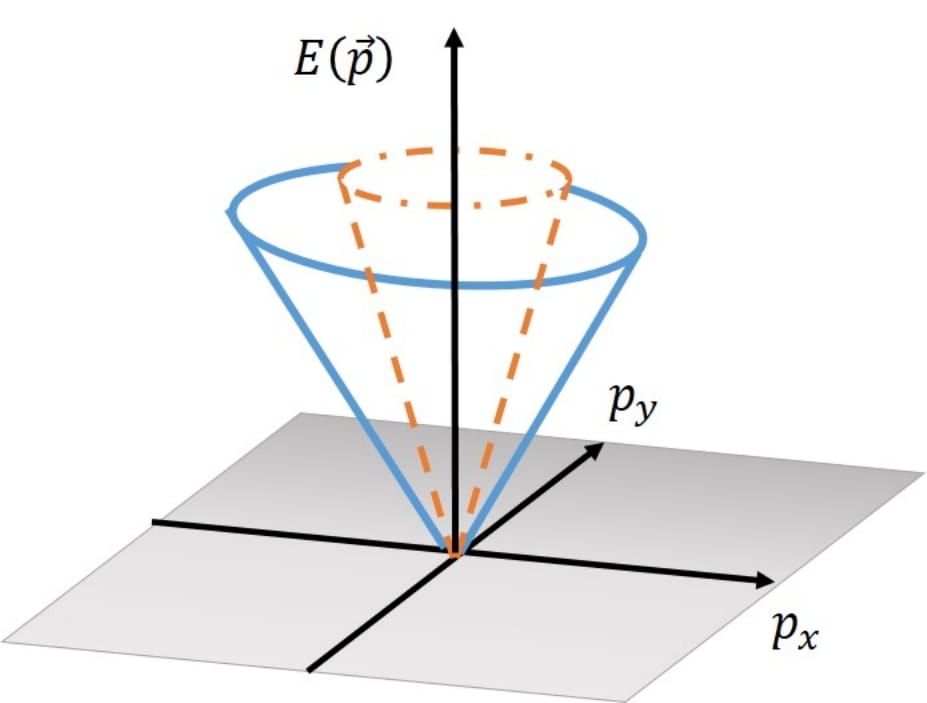}\\
		\centering{\footnotesize (a) $v_{xx}<v_{yy}$}
		\label{fig:conea}
	\end{minipage}
	\hspace{0.1cm}
	~ 
	\begin{minipage}[b]{0.45\textwidth}
		\includegraphics[width=\textwidth]{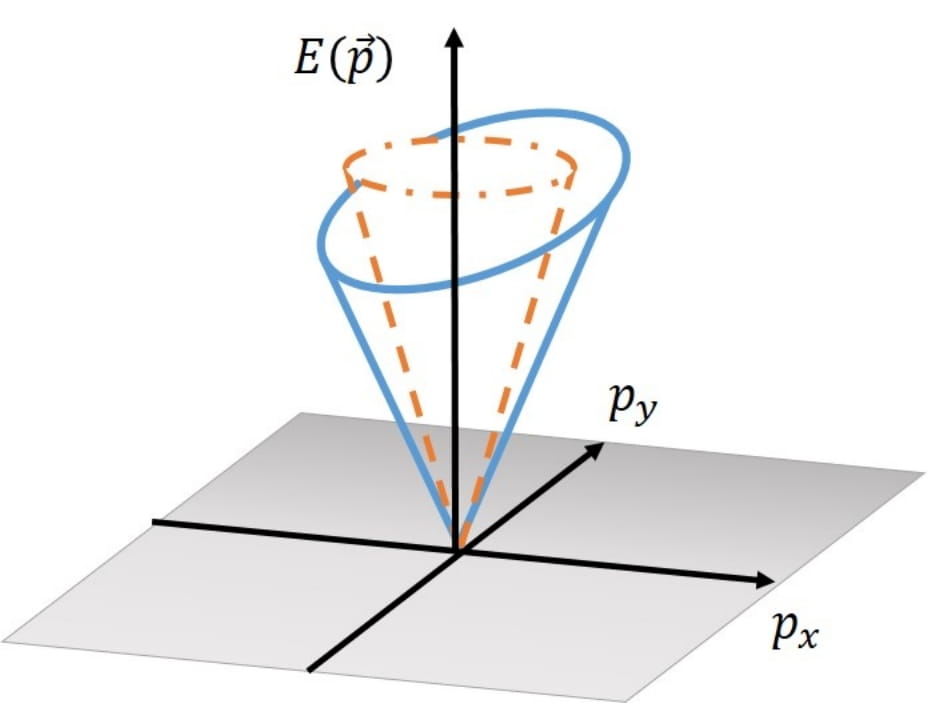}\\
		\centering{\footnotesize (b) $v_{xx}>v_{yy}$}
		\label{fig:coneb}
	\end{minipage}
	\caption{\label{fig:Dirac_cone}Dirac cones for an isotropic (dashed orange lines) and anisotropic material (solid blue lines). For former, the projections of the Dirac cones on the horizontal plane are circles, while for latter, such projections are ellipses whose semi-major axis is along either a) of the $p_x$-axis when $v_{xx}<v_{yy}$ or b) of the $p_y$-axis when $v_{xx}>v_{yy}$.}
\end{figure}

For our discussion, we adopt a particular form of the matrices $R$ and $S$ such that  the anisotropic 2D-Dirac equation~\cite{ow17,og13,pap11,tsbb15,himt16,lcwl17}, in an external magnetic field, is written as
\begin{equation}
H\Psi(x,y)=\vec{\sigma}\cdot\dvec{v}\cdot\vec{\Pi}\,\Psi(x,y)=(v_{xx}\sigma_x\pi_x+v_{yy}\sigma_y\pi_y)\Psi(x,y)=E\Psi(x,y),\label{WDE}
\end{equation}
where $\vec{\sigma}=(\sigma_x,\sigma_y)$ denotes the Pauli matrices, $\dvec{v}$ is the $2\times2$ symmetric Fermi velocity tensor with non-vanishing diagonal components $v_{xx}$ and $v_{yy}$ corresponding to the quasiparticle velocities in the directions $x$ and $y$ (see Fig.~\ref{fig:Dirac_cone}). Here, $\pi_{x,y}=p_{x,y}+eA_{x,y}/c$, with $\vec{p}$ denoting the canonical momentum and $\vec{A}$ the vector potential which defines a magnetic field aligned perpendicularly to the material surface. In a Landau-like gauge,
\begin{equation}
\vec{A}(x,y)=A_y(x)\hat{j}, \quad \vec{B}=\nabla\times\vec{A}=B(x)\hat{k},
\end{equation}
such that we can write
\begin{equation}
\Psi(x,y)=\exp(iky)\left(\begin{array}{c}
\psi^+(x) \\
\psi^-(x)
\end{array}\right).\label{spinor}
\end{equation}
Substituting~(\ref{spinor}) into~(\ref{WDE}), two coupled equations arise, namely:
\begin{equation}
\left[\sqrt{\frac{v_{xx}}{v_{yy}}}p_x\pm i\sqrt{\frac{v_{yy}}{v_{xx}}}\left(k\hbar+\frac{e}{c}A_y(x)\right)\right]\psi^\pm(x)=\frac{E}{\sqrt{v_{xx}v_{yy}}}\psi^\mp(x),
\end{equation}
These equations are decoupled to obtain
\begin{equation}
\left[-\frac{d^2}{dx^2}+V^\pm_\zeta(x)\right]\psi^\pm(x)=\epsilon^{\pm2}\psi^\pm(x), 
\end{equation}
where $\epsilon^\pm=E/v_{xx}\hbar$ and
\begin{equation}\label{9}
V^\pm_\zeta(x)=\left(\frac{k}{\zeta}+\frac{eA_y(x)}{c\hbar\,\zeta}\right)^2\pm\frac{e}{c\hbar\zeta}\frac{dA_y(x)}{dx}, \quad \zeta=\frac{v_{xx}}{v_{yy}}.
\end{equation}

In order to describe a uniform magnetic field, we take
\begin{equation}
\vec{A}=B_0x\hat{j}, \quad \vec{B}=B_0\hat{k}.
\end{equation}
Thus, by defining the frequency $\omega_{\zeta}$ as
\begin{equation}
\omega_{\zeta}=\frac{\omega_{B}}{\zeta}=\frac{2eB_0}{c\hbar\,\zeta},
\end{equation}
where $\omega_{B}$ is the cyclotron frequency of electrons in a pristine sample, we get the following Hamiltonians $H^\pm_\zeta$:
\begin{equation}\label{13}
H^\pm_\zeta=-\frac{d^2}{dx^2}+V^\pm_\zeta(x), \quad V^\pm_\zeta(x)=\frac{\omega_{\zeta}}{4}\left(x+\frac{2k}{\omega_{B}}\right)^2\pm\frac{1}{2}\omega_{\zeta}.
\end{equation}

It follows that:
\begin{equation}
\epsilon^-_{0}=0, \quad \epsilon^-_{n}=\epsilon^+_{n-1}=\omega_{\zeta}\,n, \quad n=0,1,2,\dots,
\end{equation}
or, equivalently,
\begin{equation}
E^-_{0}=0, \quad E^-_{n}=E^+_{n-1}=\hbar\sqrt{v_{xx}v_{yy}\,\omega_{B}\,n}, \quad n=0,1,2,\dots.
\end{equation}

Finally, the corresponding normalized eigenfunctions are given by:
\begin{equation}
\psi^\pm_n(x)=\sqrt{\frac{1}{2^nn!}\left(\frac{\omega_{\zeta}}{2\pi}\right)^{1/2}}\exp\left[-\frac{\omega_{\zeta}}{4}\left(x+\frac{2k}{\omega_{B}}\right)^2\right]H_n\left[\sqrt{\frac{\omega_{\zeta}}{2}}\left(x+\frac{2k}{\omega_{B}}\right)\right].
\end{equation}

\begin{figure}[h!]
	\centering
	\begin{minipage}[b]{0.45\textwidth}
		\includegraphics[width=\textwidth]{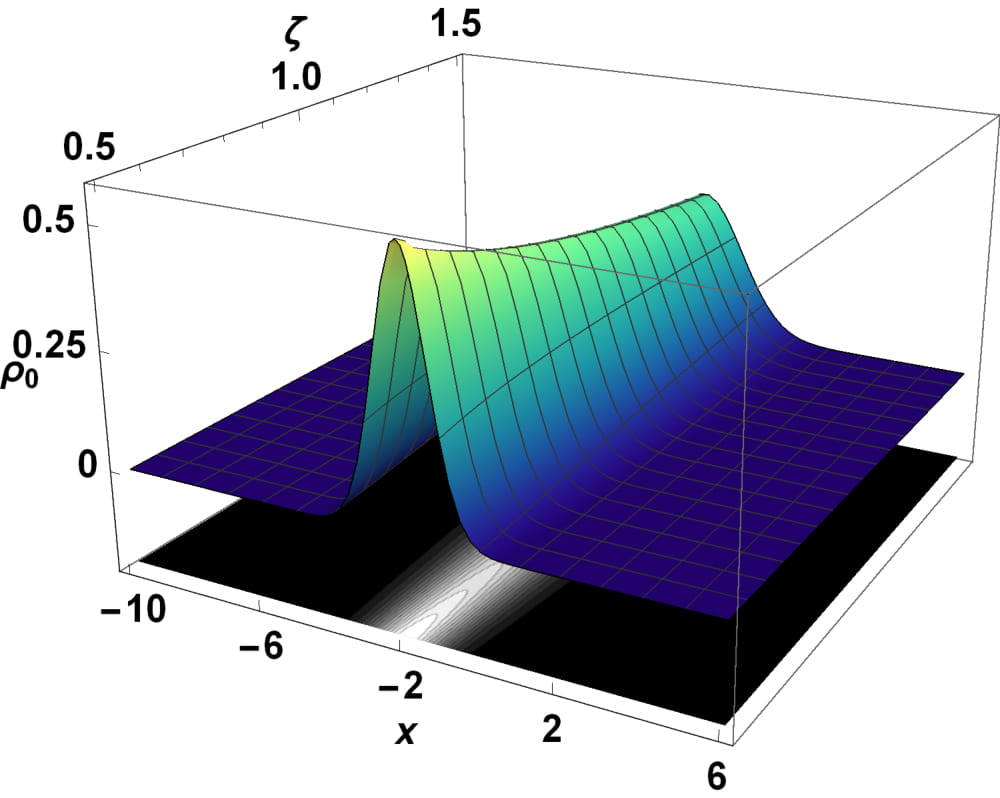}\\
		\centering{\footnotesize (a) $n=0$.}
		\label{fig:rhoN1}
	\end{minipage}
	\hspace{1cm}
	~ 
	\begin{minipage}[b]{0.45\textwidth}
		\includegraphics[width=\textwidth]{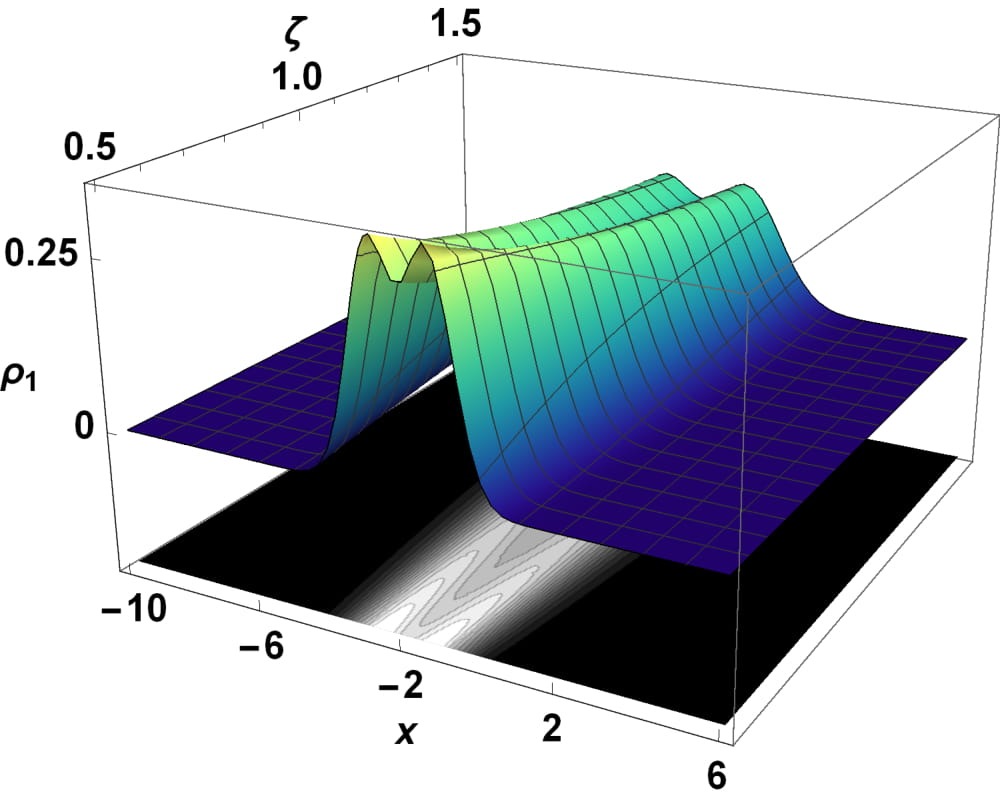}\\
		\centering{\footnotesize (b) $n=1$.}
		\label{fig:rhoN2}
	\end{minipage}
	\hspace{1cm}
	~ 
	\begin{minipage}[b]{0.45\textwidth}
		\includegraphics[width=\textwidth]{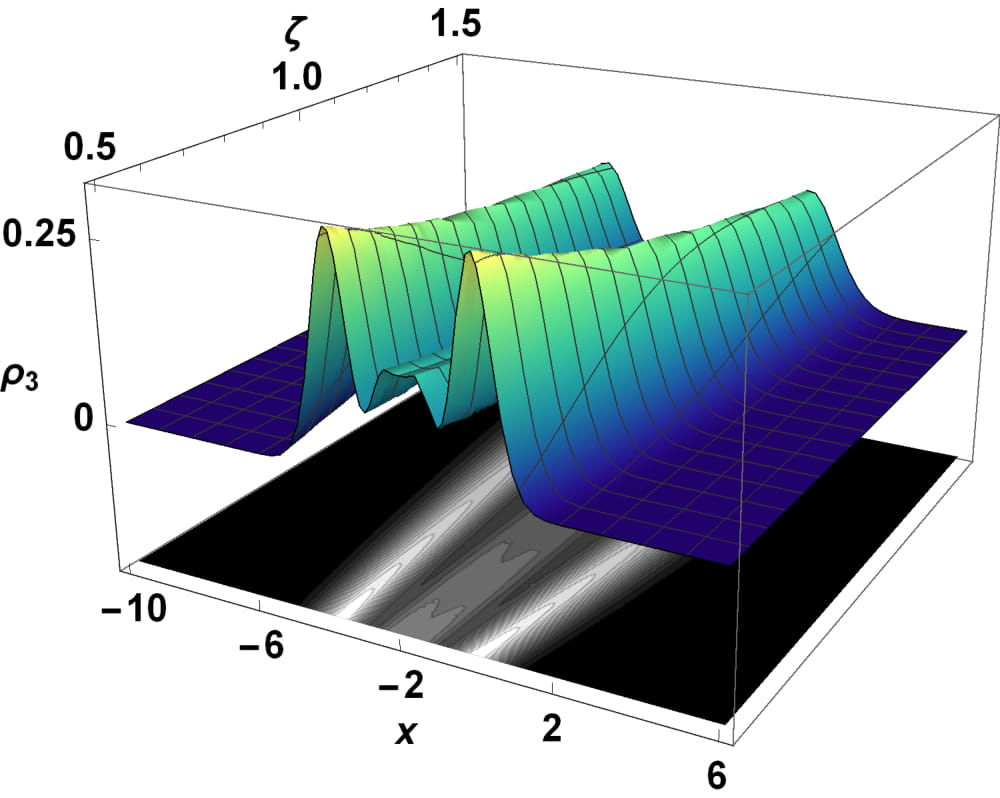}\\
		\centering{\footnotesize (c) $n=3$.}
		\label{fig:rhoN3}
	\end{minipage}
	\hspace{1cm}
	~ 
	\begin{minipage}[b]{0.45\textwidth}
		\includegraphics[width=\textwidth]{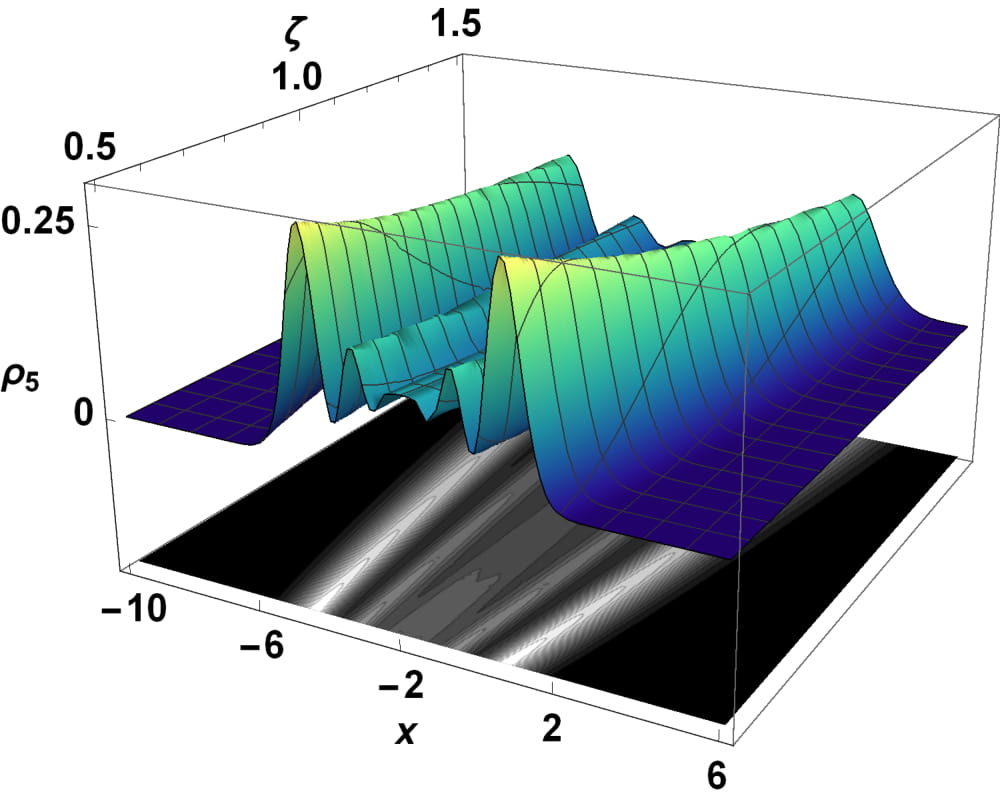}\\
		\centering{\footnotesize (d) $n=5$.}
		\label{fig:rhoN4}
	\end{minipage}
	\caption{\label{fig:rhoN}Probability density $\rho_{n}(x)$ for the pseudo-spinor states $\Psi_{n}(x,y)$ in Eq.~(\ref{17}) as function of the parameter $\zeta$ for different values of $n$. In these cases, we take $B_0=1/2,\,k=\omega_{B}=1$ and $1/2\leq\zeta\leq3/2$.}
\end{figure}

\noindent Thus, the pseudo-spinor eigenstates are
\begin{equation}\label{17}
\Psi_n(x,y)=\frac{\exp\left(iky\right)}{\sqrt{2^{(1-\delta_{0n})}}}\left(\begin{array}{c}
(1-\delta_{0n})\psi_{n-1}(x) \\
i\psi_{n}(x)
\end{array}\right), \quad n=0,1,2,\dots,
\end{equation}
where $\delta_{mn}$ denotes the Kronecker delta, $\psi^-_n\equiv\psi_{n}$ and $\psi^+_n\equiv\psi_{n-1}$.

\noindent Hence, Figure \ref{fig:rhoN} reveals two interesting facts. First, the probability density $\rho_{n}(x)$ given by
\begin{equation}
\rho_{n}(x)=\frac{1}{2^{(1-\delta_{0n})}}\left[\vert\psi_n(x)\vert^2+(1-\delta_{0n})\vert\psi_{n-1}(x)\vert^2\right],
\end{equation}
shows two maxima values in two different positions
\begin{equation}
x_{\pm}=x_{0}\pm\sqrt{\frac{2}{\omega_{\zeta}}}\,\eta,
\end{equation}
where $x_0=-2k/\omega_{B}$ and $\eta$ fulfills the polynomial relation:
\begin{equation}
g_{n}(\eta)+(1-\delta_{0n})n\,g_{n-1}(\eta)=0, \quad  g_{n}(\eta)=H_{n}(\eta)\left[\eta\,H_{n}(\eta)-H_{n+1}(\eta))\right].
\end{equation}
The distance between the points $x_{\pm}$ increases as $n$ and $\zeta$ do. In particular, we have that $x_{\pm}=x_0$ for $n=0$. Second, for given $n$ and small $\zeta$-values, the  function $\rho_{n}(x)$ takes larger values at the points $x_{\pm}$, while for growing $\zeta$-values, $\rho_{n}(x)$ takes values close to zero. In other words, if $v_{yy}>v_{xx}$, the probability to find the electron around the points $x_{\pm}$ increases while the distance respect to $x_0$ decreases. If $v_{xx}>v_{yy}$, we have the opposite situation. We deepen in this fact later on.

\subsection{Algebraic structure}
Now, let us define the following dimensionless differential operators
\begin{equation}
\theta^\pm=\frac{1}{\sqrt{2}}\left(\mp\frac{d}{d\xi}+\xi\right),\quad \theta^+=\left(\theta^-\right)^\dagger, \quad \xi=\sqrt{\frac{\omega_{\zeta}}{2}}\left(x+\frac{2k}{\omega_{B}}\right),
\end{equation}
that satisfy the commutation relation
\begin{equation}
[\theta^-,\theta^+]=1.
\end{equation}
This relation implies that the set of operators $\{\theta^+,\theta^-,1\}$ generate a HW algebra. A more general expression for the above ladder operators is discussed in \cite{ow17}.

Now, the action of the operators $\theta^\pm$ on the eigenfunctions $\psi_n$ is:
\begin{equation}
\theta^-\psi_n=\sqrt{n}\psi_{n-1}, \quad \theta^+\psi_n=\sqrt{n+1}\psi_{n+1},
\end{equation}
so that $\theta^-$ ($\theta^+$) is the annihilation (creation) operator.

In terms of these ladder operators, we can define the following dimensionless Hamiltonian $\mathcal{H}_{D}$
\begin{equation}
\mathcal{H}_{D}=\left[\begin{array}{c c}
0 & -i\theta^- \\
i\theta^+ & 0
\end{array}\right],
\end{equation}
that acts on the $x$-dependent pseudo-spinors
\begin{equation}
\Psi_n(x)=\frac{1}{\sqrt{2^{(1-\delta_{0n})}}}\left(\begin{array}{c}
(1-\delta_{0n})\psi_{n-1}(x) \\
i\psi_{n}(x)
\end{array}\right), \quad n=0,1,2,\dots.
\end{equation}

\section{Annhilation operator}\label{annihilation}
In order to build nonlinear coherent states in 2D-DM, one can define a deformed annihilation operator $\Theta_f$ given by:
\begin{equation}
\Theta_f^-=\left[\begin{array}{c c}
\cos(\delta)\frac{\sqrt{N+2}}{\sqrt{N+1}}f(N+2)\theta^- & \sin(\delta)\frac{f(N+2)}{\sqrt{N+1}}(\theta^-)^2 \\
-\sin(\delta)f(N+1)\sqrt{N+1} & \cos(\delta)f(N+1)\theta^-
\end{array}\right], \quad \Theta_f^+=(\Theta_f^-)^\dagger,
\end{equation}
such that
\begin{equation}
\Theta_f^-\Psi_{n}(x,y)=\frac{f(n)}{\sqrt{2^{\delta_{1n}}}}\exp(i\delta)\sqrt{n}\Psi_{n-1}(x,y), \quad n=0,1,2,\dots,
\end{equation}
where $f(N)$ is again a well-behaved function of the number operator $N=\theta^+\theta^-$ and $\delta\in[0,2\pi]$ is a parameter that allows us to consider either diagonal or non-diagonal matrix representation for $\Theta_f^\pm$. Also, these operators satisfy the nonlinear algebra
\begin{equation}
[\Theta_f^-,\Theta_f^+]=\left[\begin{array}{c c}
\Omega(N+1) & 0 \\
0 & \Omega(N)
\end{array}\right], \quad \Omega(N)=(N+1)f^2(N+1)-Nf^2(N).
\end{equation}
In the limit $f(N)=1$, we have that $[\Theta_f^-,\Theta_f^+]=\mathbb{I}$, where $\mathbb{I}$ is the $2\times2$ unity matrix, {\it i.e.}, we recover the HW algebra.

\subsection{Nonlinear coherent states}
We can construct NLCS $\Psi^f_{\alpha}(x,y)$ as eigenstates of the operator $\Theta_f^-$:
\begin{equation}\label{31}
\Theta_f^-\Psi^f_{\alpha}(x,y)=\alpha\Psi^f_{\alpha}(x,y), \quad \alpha\in\mathbb{C},
\end{equation}
where
\begin{equation}
\Psi^f_{\alpha}(x,y)=a_0\Psi_{0}(x,y)+\sum_{n=1}^{\infty}a_n\Psi_{n}(x,y).
\end{equation}

Upon inserting these states into the corresponding eigenvalue equation~(\ref{31}), we get the following relations:
\begin{equation}
a_1f(1)=\sqrt{2}\tilde{\alpha}a_0, \quad a_{n+1}f(n+1)\sqrt{n+1}=\tilde{\alpha}a_n,
\end{equation}
with $\tilde{\alpha}=\alpha\exp(-i\delta)$. This means that to work with either a diagonal or non-diagonal annihilation operator $\Theta_f^-$ results in the introduction of a phase factor that affects the eigenvalue $\alpha$.

From here, the construction of the nonlinear coherent states is identical to one discussed in \cite{ed17}, along the same cases according to the function $f(N)$. Thus, we focus in giving some examples of such states in order to describe the effects of strain on the NLCS.

%
%
%

\subsection{Some examples}
It is worth to mention that in the discussion above, one can choose any form for the function $f(N)$ that characterizes the NLCS provided that it retains the  convergence of the series involved and hence guaranties that such coherent states still belong to the Hilbert space. However, depending on such a function $f(N)$, one would have the possibility to introduce a different description from the harmonic oscillator to get a {\it deformed} dynamics in phase space \cite{mmzs96,mmsz97}. Therefore, in order to describe the effects of strain on the NLCS, in the following sections we  consider some particular forms for the function $f(N+1)$ in $\Theta_f^-$ \cite{ed17}. Moreover, we make use of use some physical quantities to analyze such quantum states, including the probability density $\rho_{\alpha}(x)$, the mean energy $\langle H\rangle$, the occupation number distribution $P_{\alpha}(n)$ and the Heisenberg uncertainty relation (HUR). To compute the latter, we define the matrix operator $\mathbb{S}_q$ and its square as
\begin{equation}
\mathbb{S}_q=s_q\otimes\mathbb{I}, \quad \mathbb{S}_q^2=s_q^2\otimes\mathbb{I},
\end{equation}
where
\begin{subequations}
	\begin{eqnarray}
	&&s_q=\frac{1}{\sqrt{2}i^q}\left(\theta^-+(-1)^q\theta^+\right), \\
	&&s_q^2=\frac{1}{2}\left[2N+1+(-1)^q((\theta^-)^2+(\theta^+)^2)\right],
	\end{eqnarray}
\end{subequations}
and $q=0,1$. The variance of the operator $\mathbb{S}_q$ is calculated as follows:
\begin{equation}
\sigma_{\mathbb\mathbb{S}_q}=\sqrt{\langle\mathbb{S}_q^2\rangle-\langle\mathbb{S}_q\rangle^2}.
\end{equation}
Thus, when $q=0$ ($q=1$), we have that $\sigma_{\mathbb\mathbb{S}_0}\equiv\sigma_{\xi}$ ($\sigma_{\mathbb\mathbb{S}_1}\equiv\sigma_{p}$), {\it i.e.}, the variance of the position $\xi$ (momentum $p$) operator and the HUR must fulfill:
\begin{equation}
\sigma_{\xi}\sigma_{p}=\sigma_{\mathbb\mathbb{S}_0}\sigma_{\mathbb\mathbb{S}_1}\geq\frac12.
\end{equation}

\subsubsection{Case for $f(1)\neq0$}
\begin{figure}[h!]
	\centering
	\begin{minipage}[b]{0.45\textwidth}
		\includegraphics[width=\textwidth]{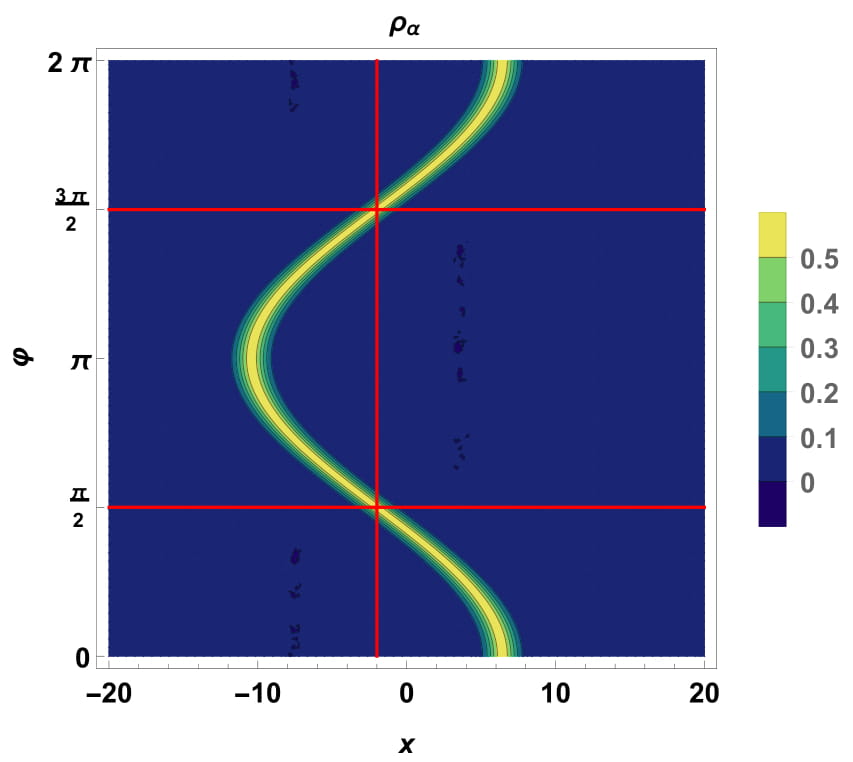}\\
		\centering{\footnotesize (a) $\zeta=1/2$.}
		\label{fig:rhoIA}
	\end{minipage}
	\hspace{1cm}
	~ 
	\begin{minipage}[b]{0.45\textwidth}
		\includegraphics[width=\textwidth]{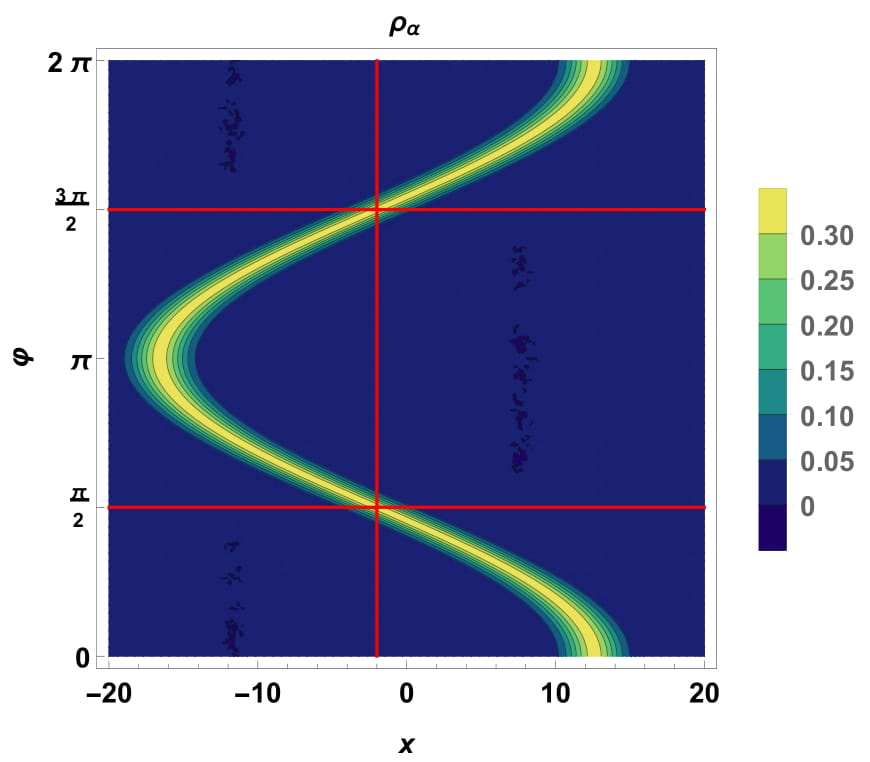}\\
		\centering{\footnotesize (b) $\zeta=3/2$.}
		\label{fig:rhoIB}
	\end{minipage}
	\caption{\label{fig:rhoIa}Probability density $\rho_{\alpha}(x)$ of the coherent states $\Psi^f_{\alpha}(x,y)$ in Eq.~(\ref{37}) for $\vert\alpha\vert=6$ and some values of the parameter $\zeta$. In these cases, we take $B_0=1/2$, $k=\omega_{B}=1$ and $\delta=0$.}
\end{figure}

\begin{figure}[h!]
	\centering
	\includegraphics[width=0.7\textwidth]{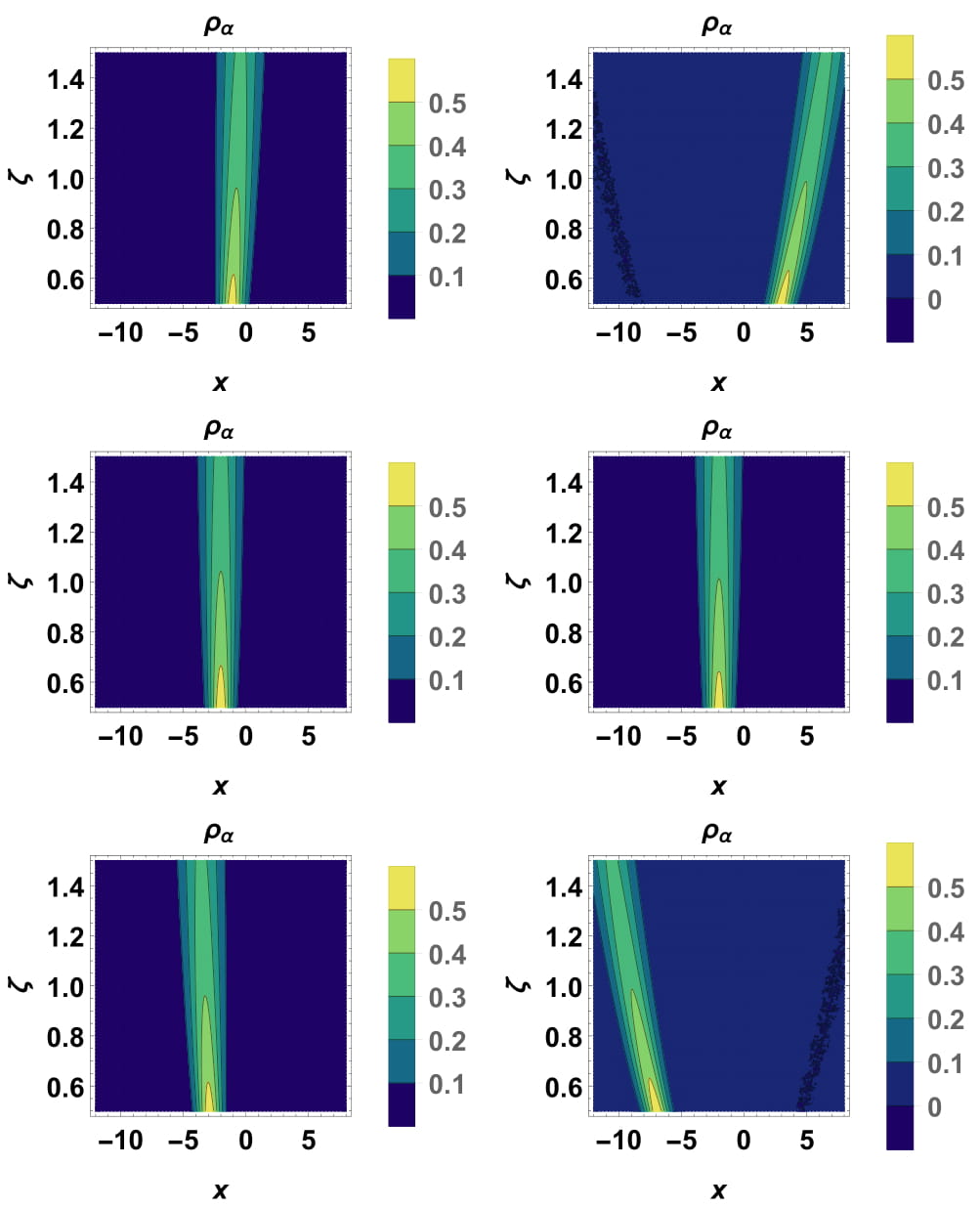}
	\caption{\label{fig:rhoI}Probability density $\rho_{\alpha}(x)$ for the coherent states $\Psi^f_{\alpha}(x,y)$ in Eq.~(\ref{37}) as function of the parameter $\zeta$ for different values of eigenvalue $\alpha=\vert\alpha\vert\exp\left(i\varphi\right)$: (vertical) $\vert\alpha\vert=1,5$, and (horizontal) $\varphi=\pi/4,\pi/2,3\pi/4$. In all these cases, we take $B_0=1/2$, $k=\omega_{B}=1$ and $\delta=0$.}
\end{figure}

The simplest form for $f(N)$ that satisfies the condition $f(1)\neq0$ is $f(N+1)=1$. For this choice, the corresponding NLCS are given by
\begin{equation}\label{37}
\Psi_{\alpha}^f(x,y)=\frac{1}{\sqrt{2\exp\left(\vert\tilde{\alpha}\vert^2\right)-1}}\left[\Psi_{0}(x,y)+\sum_{n=1}^{\infty}\frac{\sqrt{2}\,\tilde{\alpha}^n}{\sqrt{n!}}\Psi_{n}(x,y)\right],
\end{equation}
whose probability density is depicted in Figs.~\ref{fig:rhoIa} and \ref{fig:rhoI} and has the analytical form:
\begin{eqnarray}
\nonumber\rho_{\alpha}(x)&=&\Psi_{\alpha}^f(x,y)^\dagger\Psi_{\alpha}^f(x,y)=\frac{1}{2\exp\left(\vert\tilde{\alpha}\vert^2\right)-1}\left[\psi_{0}^2(x)+\left\vert\sum\limits_{n=1}^{\infty}\frac{\tilde{\alpha}^n}{\sqrt{n!}}\psi_{n}(x)\right\vert^2\right.\nonumber\\
&&\left.+\left\vert\sum\limits_{n=1}^{\infty}\frac{\tilde{\alpha}^n}{\sqrt{n!}}\psi_{n-1}(x)\right\vert^2+2\Re\left(\sum\limits_{n=1}^{\infty}\frac{\tilde{\alpha}^n}{\sqrt{n!}}\psi_{n}(x)\psi_{0}(x)\right)\right].
\end{eqnarray}

\begin{figure}[h!]
	\centering
	\begin{minipage}[b]{0.48\textwidth}
		\includegraphics[width=\textwidth]{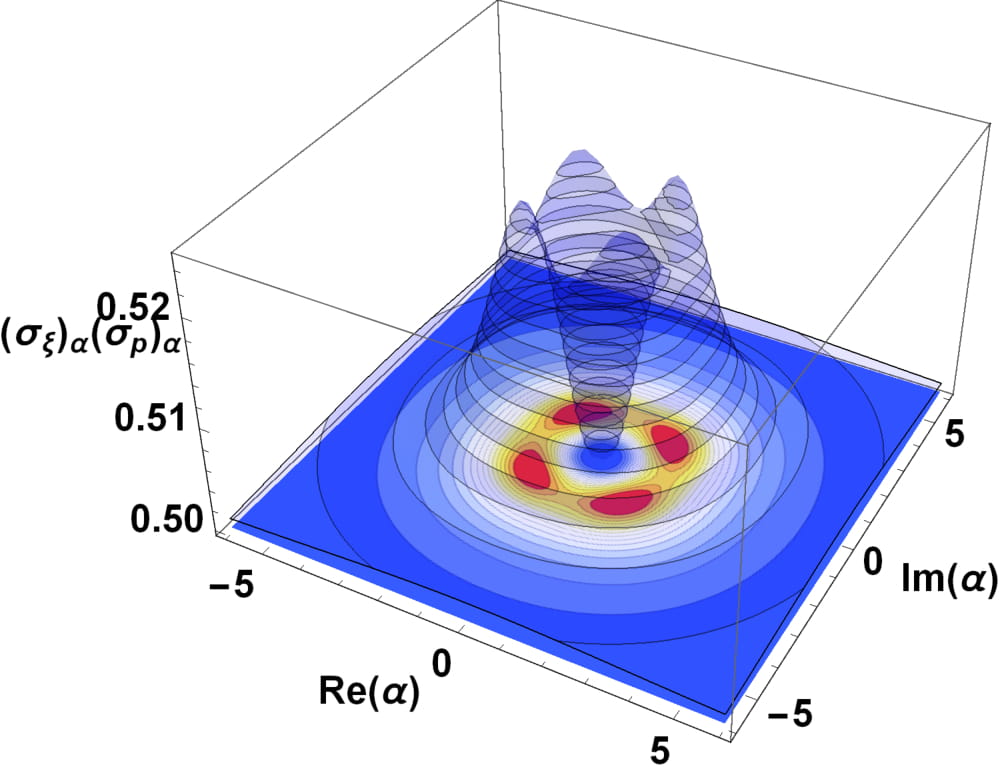}\\
		\centering{\footnotesize (a) $(\sigma_\xi)_\alpha(\sigma_p)_\alpha$.}
		\label{fig:HUR1}
	\end{minipage}
	\hspace{0.7cm}
	~ 
	\begin{minipage}[b]{0.41\textwidth}
		\includegraphics[width=\textwidth]{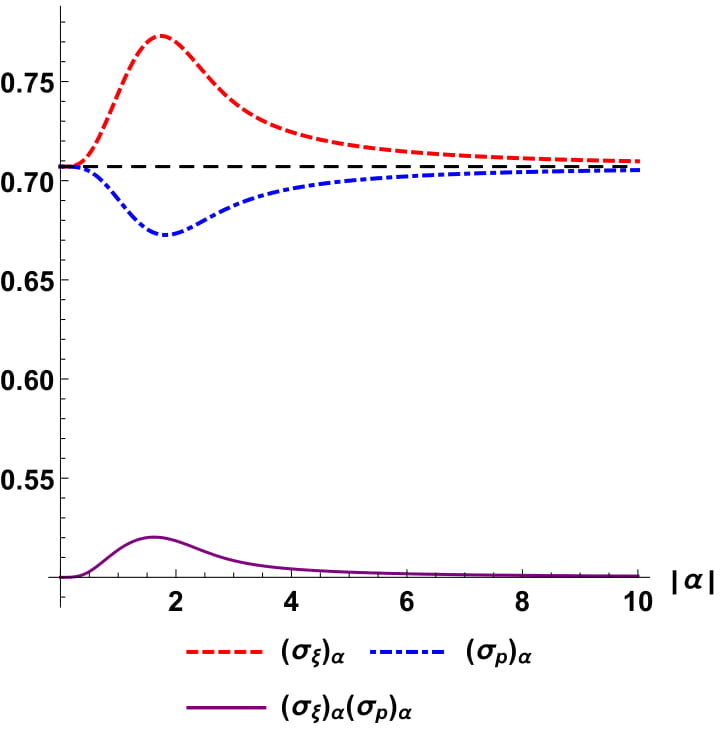}\\
		\centering{\footnotesize (b) $\varphi=0$.}
		\label{fig:HUR2}
	\end{minipage}
	\hspace{1cm}
	~ 
	\begin{minipage}[b]{0.41\textwidth}
		\includegraphics[width=\textwidth]{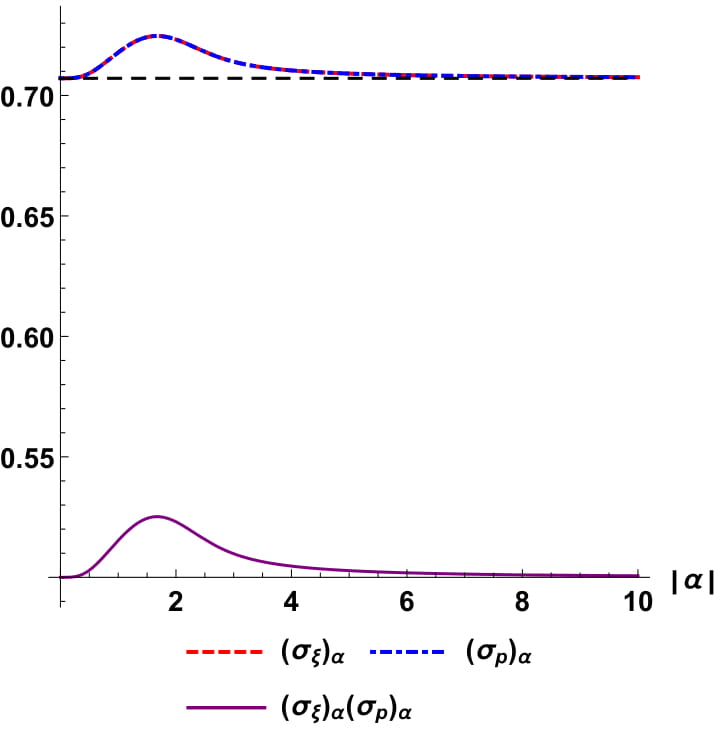}\\
		\centering{\footnotesize (c) $\varphi=\pi/4$.}
		\label{fig:HUR3}
	\end{minipage}
	\hspace{1.5cm}
	~ 
	\begin{minipage}[b]{0.41\textwidth}
		\includegraphics[width=\textwidth]{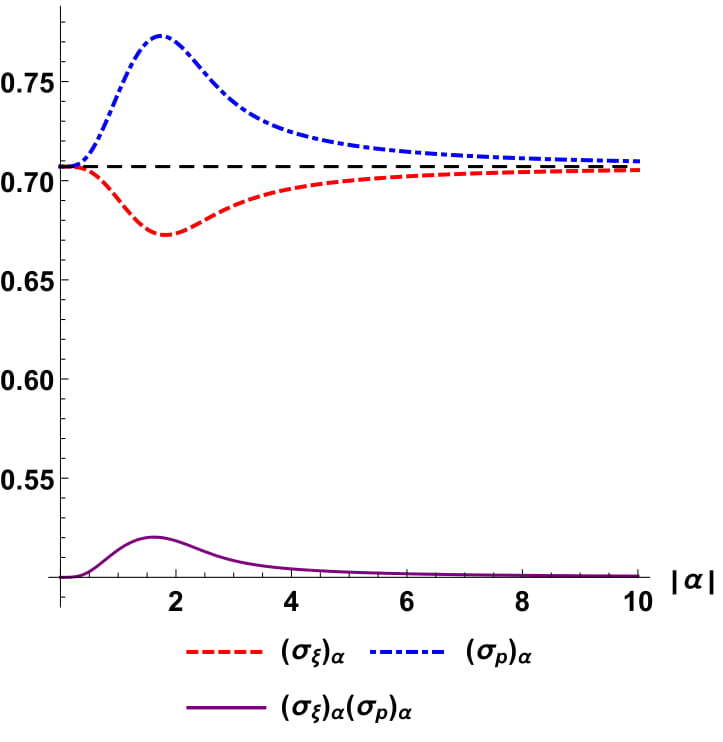}\\
		\centering{\footnotesize (d) $\varphi=\pi/2$.}
		\label{fig:HUR4}
	\end{minipage}
	\caption{\label{fig:HUR}For the states in Eq.~(\ref{37}): (a) $(\sigma_\xi)_\alpha(\sigma_p)_\alpha$ as function of $\alpha$.~(b-d) Comparison between $(\sigma_\xi)_\alpha$, $(\sigma_p)_\alpha$ and $(\sigma_\xi)_\alpha(\sigma_p)_\alpha$ as function of $\vert\alpha\vert$. As $\vert\alpha\vert$ increases both $(\sigma_\xi)_\alpha$ and $(\sigma_p)_\alpha$ approach the value $1/\sqrt{2}$ and thus their product tends to the value $1/2$. Also, as  $\varphi$ changes, the dispersion of the position $\xi$ is upper, equal or lower than that of the momentum $p$.}
\end{figure}

Using these NLCS, the mean values of the operators $\mathbb{S}_q$ and $\mathbb{S}^2_q$ are, respectively (see Fig.~\ref{fig:HUR}):
\small
\begin{subequations}
	\begin{align}
	\langle\mathbb{S}_q\rangle_{\alpha}&=\frac{\tilde{\alpha}+(-1)^q\tilde{\alpha}^\ast}{\sqrt{2}i^q(2\exp\left(\vert\tilde{\alpha}\vert^2\right)-1)}\left[\exp\left(\vert\tilde{\alpha}\vert^2\right)+\sum_{n=1}^{\infty}\frac{\vert\tilde{\alpha}\vert^{2n}}{\sqrt{(n-1)!(n+1)!}}\right], \\
	\nonumber\langle\mathbb{S}^2_q\rangle_{\alpha}&=\frac{1}{2(2\exp\left(\vert\tilde{\alpha}\vert^2\right)-1)}\Bigg[1+4\vert\tilde{\alpha}\vert^2\exp\left(\vert\tilde{\alpha}\vert^2\right)+(-1)^q(\tilde{\alpha}^2+\tilde{\alpha}^{\ast 2})\times\\
	&\quad\times\Bigg[\exp\left(\vert\tilde{\alpha}\vert^2\right)+\sum_{n=1}^{\infty}\frac{\sqrt{n+1}\,\vert\tilde{\alpha}\vert^{2n}}{\sqrt{(n-1)!(n+2)!}}\Bigg]\Bigg],
	\end{align}
\end{subequations}
\normalsize
while the mean energy $\langle H\rangle_\alpha^\zeta$ turns out to be (see Fig.~\ref{fig:Halpha_strain}):
\begin{equation}\label{41}
\langle H\rangle_\alpha^\zeta=\sqrt{v_{xx}v_{yy}}\langle H\rangle_\alpha, \quad \langle H\rangle_\alpha=\frac{2\sqrt{\omega_{B}}\,\hbar}{2\exp\left(\vert\tilde{\alpha}\vert^2\right)-1}\sum_{n=1}^{\infty}\frac{\vert\tilde{\alpha}\vert^{2n}}{n!}\sqrt{n},
\end{equation}
where $\langle H\rangle_\alpha$ is the mean energy for a pristine 2D-DM for the case $f(1)\neq0$.

In a semi-classical interpretation, the eigenvalue $\alpha=\vert\alpha\vert\exp\left(i\varphi\right)$ determines the initial conditions of the motion of the electrons. As $\vert\alpha\vert$ changes, the maximum probability density moves along the $x$-axis, {\it i.e.}, the center of  $\rho_{\alpha}(x)$  moves away from or approaches to the equilibrium position $x_0=2k/\omega_{B}$. Also, if $\varphi\in[0,2\pi]$ varies, the maximum probability performs an oscillatory-like motion around  $x_0$ (vertical red line in Fig.~\ref{fig:rhoIa}). In particular, for $\varphi=$~Arg$(\alpha)=(2m+1)\pi/2$, $m=0,1,\dots$, $\rho_{\alpha}(x)$ is located around the position $x_0$ (horizontal red lines in Fig.~\ref{fig:rhoIa}). Note that for a given eigenvalue $\alpha$ and for this one and the following NLCS, $\delta\neq0$ will allow to localize the maximum probability closer or further away from $x_0$, in relation to where the potentials $V_{\zeta}^{\pm}(x)$ in Eq.~(\ref{13}) take their minimum value and define the so-called return points, which in turn depend on $\zeta$.

On the other hand, the parameter $\zeta$ affects the value of the probability density, as shown in Fig.~\ref{fig:rhoI}. Similarily to what happens with the probability density of the spinorial eigenstates $\Psi_{n}$, the function $\rho_{\alpha}(x)$ is larger when $\zeta$ gets small, while in the opposite regime, $\zeta>1$, $\rho_{\alpha}(x)$ tends to zero. Moreover, the maximum probability density is located either to the right or to the left of the equilibrium point $x_0$ according to $0\leq\varphi<\pi/2$ or $\pi/2<\varphi\leq2\pi$. For $\varphi=\pi/2$, the center of  $\rho_{\alpha}(x)$ remains at $x_0$.

Finally, Fig.~\ref{fig:HUR} shows that the Heisenberg uncertainty relation reaches a maximum value for small values of $\vert\alpha\vert$ and $\varphi=\pi/4$, while in the limits $\alpha\rightarrow0$ and $\alpha\rightarrow\infty$ we have that $(\sigma_{\xi})_{\alpha}(\sigma_{p})_{\alpha}\rightarrow1/2$. This behavior can be understood through the respective variances of the position $\xi $ and momentum $p$ operators: when $\varphi=0$, the function $\langle\mathbb{S}_{1}\rangle_{\alpha}=\langle p\rangle_{\alpha}=0$  and the dispersion of the momentum $p$ is smaller than that of the position $\xi$. As $\varphi$ grows, the dispersions of each operator change until they are equal ($\varphi=\pi/4$) or their behaviors are exchanged ($\varphi=\pi/2$), {\it i.e.}, now we have that $\langle\mathbb{S}_{0}\rangle_{\alpha}=\langle\xi\rangle_{\alpha}=0$. This last circumstance implies that the electron performs symmetric oscillations around the equilibrium position $x_0$, in agreement to the previous analysis of the probability density.

\subsubsection{Case for $f(1)=0$}
Now, we consider the case for $f(1)=0$. As we mentioned in the previous section, we can consider two new cases.

\begin{figure}[h!]
	\centering
	\begin{minipage}[b]{0.45\textwidth}
		\includegraphics[width=\textwidth]{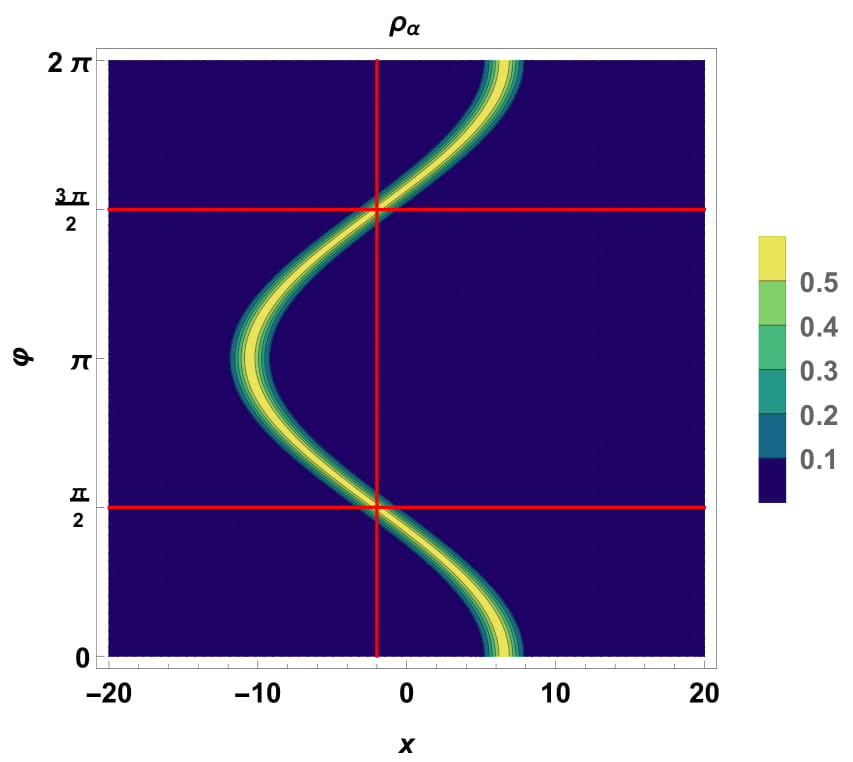}\\
		\centering{\footnotesize (a) $\zeta=1/2$.}
		\label{fig:rhoIIA}
	\end{minipage}
	\hspace{1cm}
	~ 
	\begin{minipage}[b]{0.45\textwidth}
		\includegraphics[width=\textwidth]{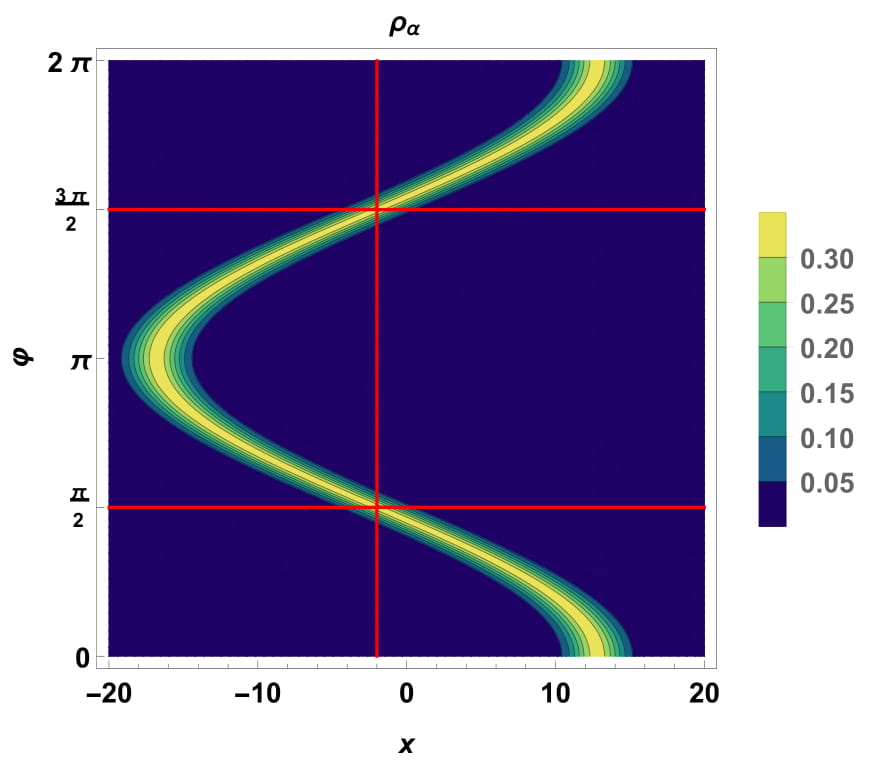}\\
		\centering{\footnotesize (b) $\zeta=3/2$.}
		\label{fig:rhoIIB}
	\end{minipage}
	\caption{\label{fig:rhoIIa}Probability density $\rho_{\alpha}(x)$ of the coherent states $\Psi^f_{\alpha}(x,y)$ in Eq.~(\ref{42}) for $\vert\alpha\vert=6$ and some values of the parameter $\zeta$. In these cases, we take $B_0=1/2$, $k=\omega_{B}=1$ and $\delta=0$.}
\end{figure}

\begin{figure}[h!]
	\centering
	\includegraphics[width=0.7\textwidth]{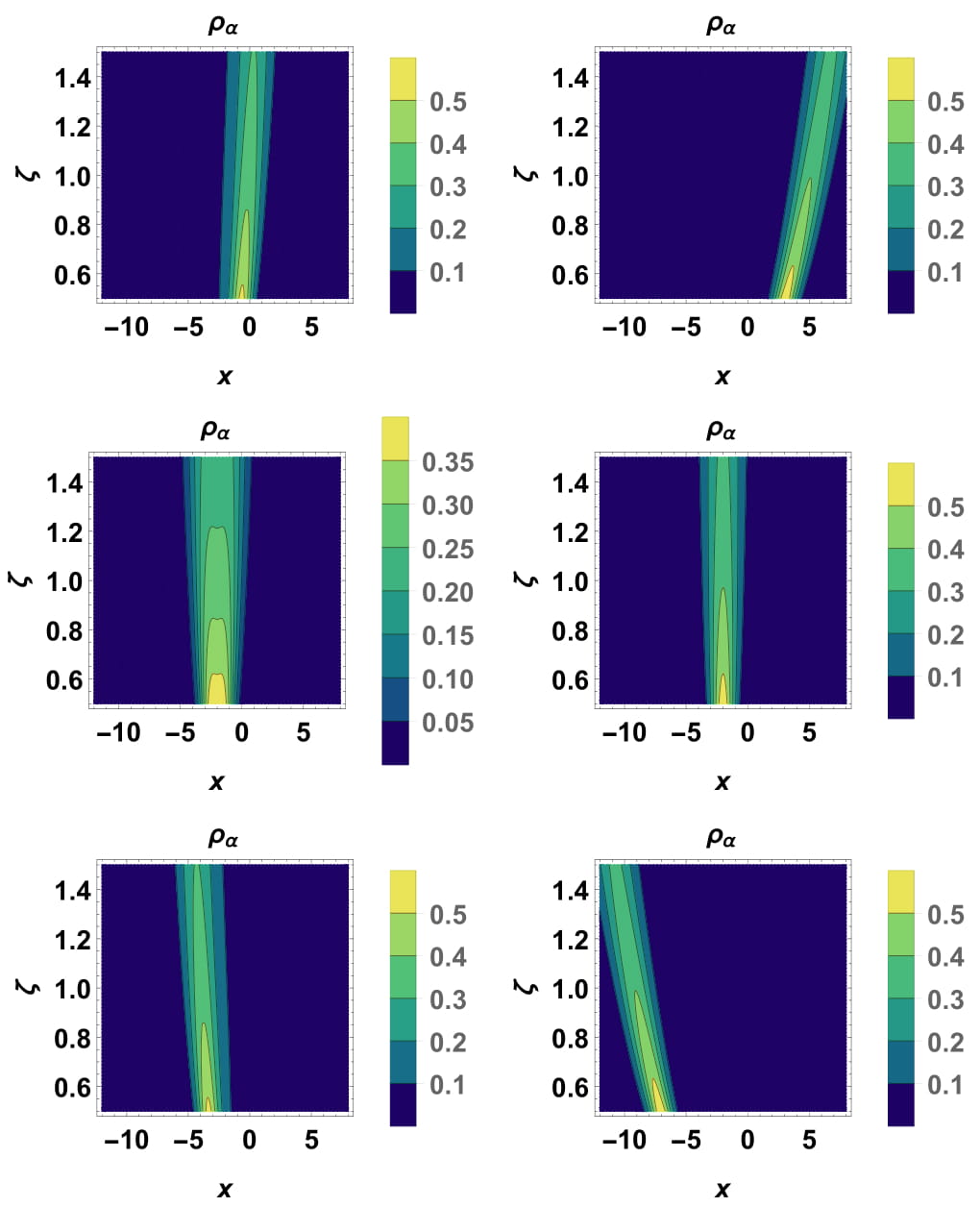}
	\caption{\label{fig:rhoII}Probability density $\rho_{\alpha}(x)$ for the coherent states $\Psi^f_{\alpha}(x,y)$ in Eq.~(\ref{42}) as function of the parameter $\zeta$ for different values of eigenvalue $\alpha=\vert\alpha\vert\exp\left(i\varphi\right)$: (vertical) $\vert\alpha\vert=1,5$, and (horizontal) $\varphi=\pi/4,\pi/2,3\pi/4$. In all these cases, we take $B_0=1/2$, $k=\omega_{B}=1$ and $\delta=0$.}
\end{figure}

\paragraph{a) $f(2)\neq0$}
A function $f(N)$ that satisfies the additional condition $f(2)\neq0$ is $f(N+1)=g(N)=\sqrt{N}/\sqrt{N+1}$. Hence, the NLCS turn out to be
\begin{equation}\label{42}
\Psi_{\alpha}^f(x,y)=\exp\left(-\frac{\vert\tilde{\alpha}\vert^2}{2}\right)\sum_{n=0}^{\infty}\frac{\tilde{\alpha}^n}{\sqrt{n!}}\Psi_{n+1}(x,y),
\end{equation}
and its probability density is (see Figs.~\ref{fig:rhoIIa} and \ref{fig:rhoII}):
\begin{eqnarray}
\rho_{\alpha}(x)&=&\Psi_{\alpha}^f(x,y)^\dagger\Psi_{\alpha}^f(x,y)=\frac{\exp\left(-\vert\tilde{\alpha}\vert^2\right)}{2}\left[\left\vert\sum\limits_{n=0}^{\infty}\frac{\tilde{\alpha}^n}{\sqrt{n!}}\psi_{n+1}(x)\right\vert^2+\left\vert\sum\limits_{n=0}^{\infty}\frac{\tilde{\alpha}^n}{\sqrt{n!}}\psi_{n}(x)\right\vert^2\right].\label{45}
\end{eqnarray}

\begin{figure}[h!]
	\centering
	\begin{minipage}[b]{0.48\textwidth}
		\includegraphics[width=\textwidth]{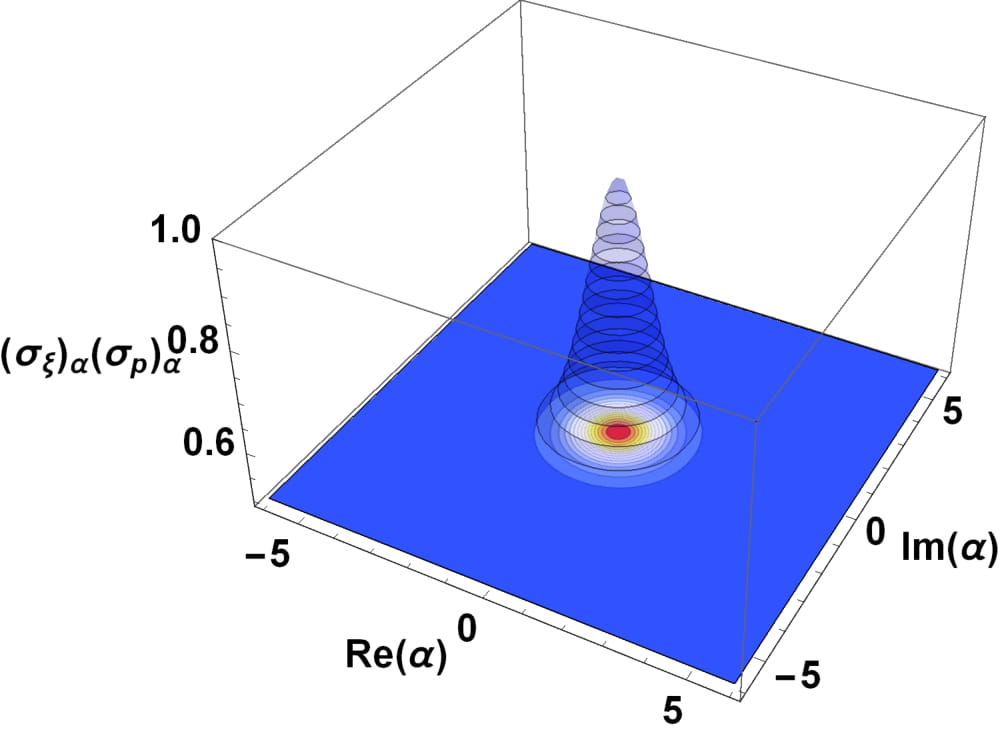}\\
		\centering{\footnotesize (a) $(\sigma_\xi)_\alpha(\sigma_p)_\alpha$.}
		\label{fig:HURII1}
	\end{minipage}
	\hspace{0.7cm}
	~ 
	\begin{minipage}[b]{0.41\textwidth}
		\includegraphics[width=\textwidth]{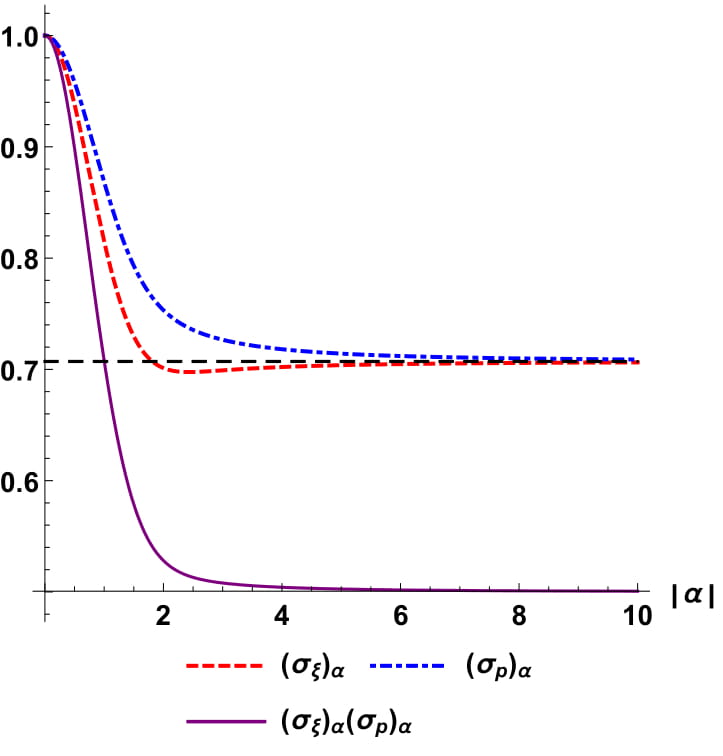}\\
		\centering{\footnotesize (b) $\varphi=0$.}
		\label{fig:HURII2}
	\end{minipage}
	\hspace{1cm}
	~ 
	\begin{minipage}[b]{0.41\textwidth}
		\includegraphics[width=\textwidth]{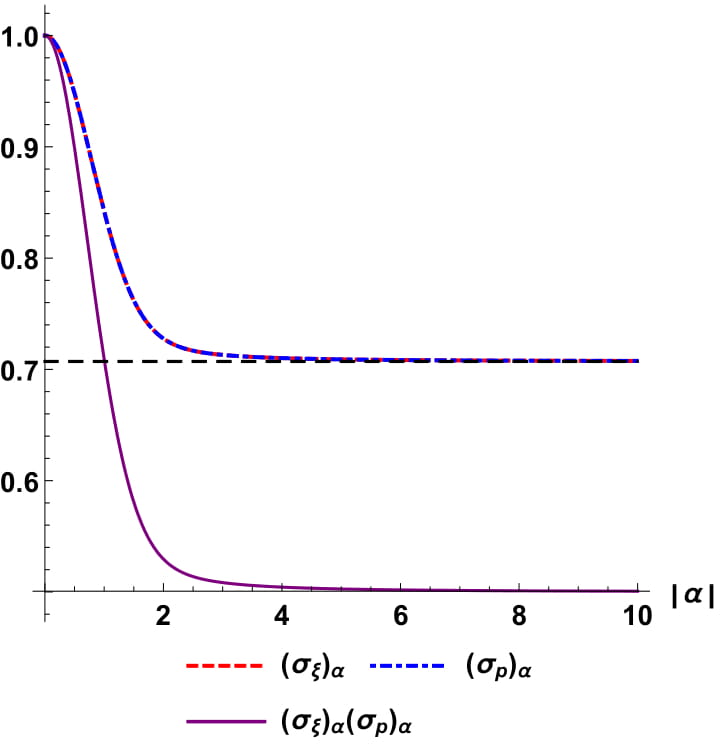}\\
		\centering{\footnotesize (c) $\varphi=\pi/4$.}
		\label{fig:HURII3}
	\end{minipage}
	\hspace{1.5cm}
	~ 
	\begin{minipage}[b]{0.41\textwidth}
		\includegraphics[width=\textwidth]{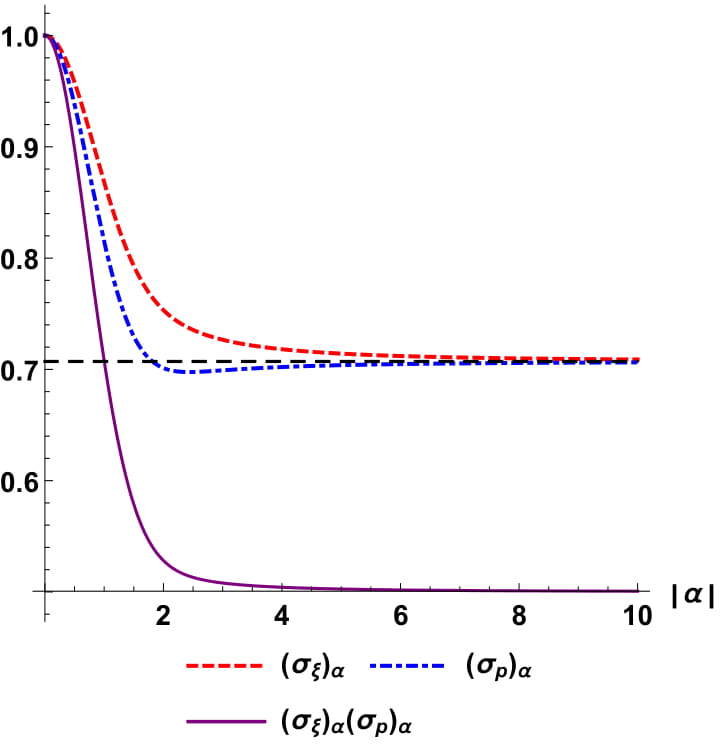}\\
		\centering{\footnotesize (d) $\varphi=\pi/2$.}
		\label{fig:HURII4}
	\end{minipage}
	\caption{\label{fig:HURII}For the states in Eq.~(\ref{42}): (a) $(\sigma_\xi)_\alpha(\sigma_p)_\alpha$ as function of $\alpha$.~(b-d) Comparison between $(\sigma_\xi)_\alpha$, $(\sigma_p)_\alpha$ and $(\sigma_\xi)_\alpha(\sigma_p)_\alpha$ as function of $\vert\alpha\vert$. As $\vert\alpha\vert$ increases both $(\sigma_\xi)_\alpha$ and $(\sigma_p)_\alpha$ approach the value $1/\sqrt{2}$ and thus their product tends to $1/2$. Also, as  $\varphi$ changes, the dispersion of the position $\xi$ is lower, equal or upper than that of the momentum $p$.}
\end{figure}

The mean values of the operators $\mathbb{S}_q$ and $\mathbb{S}^2_q$ in this representation are, respectively (see Fig.~\ref{fig:HURII}):
\begin{subequations}
	\begin{align}
	\langle\mathbb{S}_q\rangle_{\alpha}&=\frac{\tilde{\alpha}+(-1)^q\tilde{\alpha}^\ast}{2\sqrt{2}i^q}\left[1+\exp\left(-\vert\tilde{\alpha}\vert^2\right)\sum_{n=0}^{\infty}\frac{\sqrt{n+2}\,\vert\tilde{\alpha}\vert^{2n}}{\sqrt{n!(n+1)!}}\right], 
	\end{align}
	\begin{align}
	\langle\mathbb{S}^2_q\rangle_{\alpha}&=1+\vert\tilde{\alpha}\vert^2+(-1)^q\frac{(\tilde{\alpha}^2+\tilde{\alpha}^{\ast 2})}{4}\left[1+\exp\left(-\vert\tilde{\alpha}\vert^2\right)\sum_{n=0}^{\infty}\frac{\sqrt{n+3}\,\vert\tilde{\alpha}\vert^{2n}}{\sqrt{n!(n+1)!}}\right],
	\end{align}
\end{subequations}


\begin{figure}[h!]
	\centering
	\begin{minipage}[b]{0.45\textwidth}
		\includegraphics[width=\textwidth]{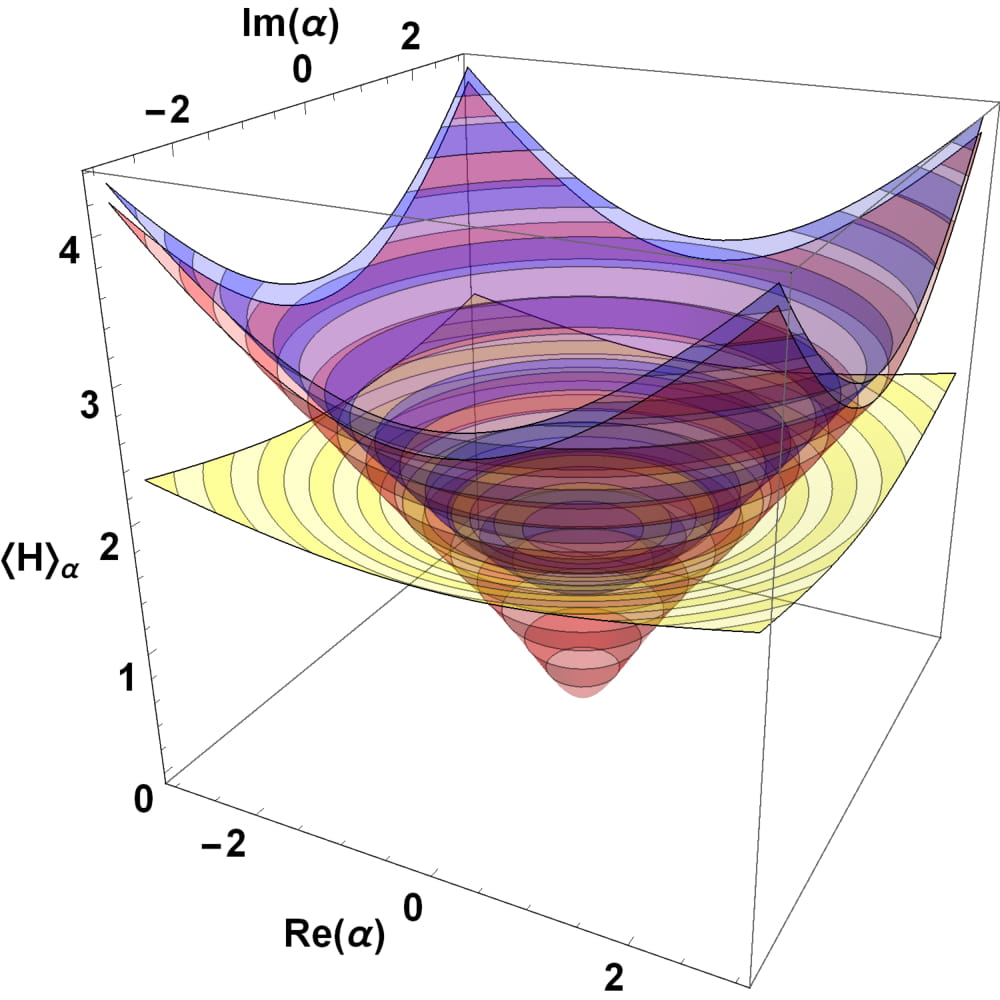}\\
		\centering{\footnotesize (a) $B_{0}=1/2$.}
		\label{fig::Halpha_strain_a}
	\end{minipage}
	\hspace{0.7cm}
	~ 
	\begin{minipage}[b]{0.47\textwidth}
		\includegraphics[width=\textwidth]{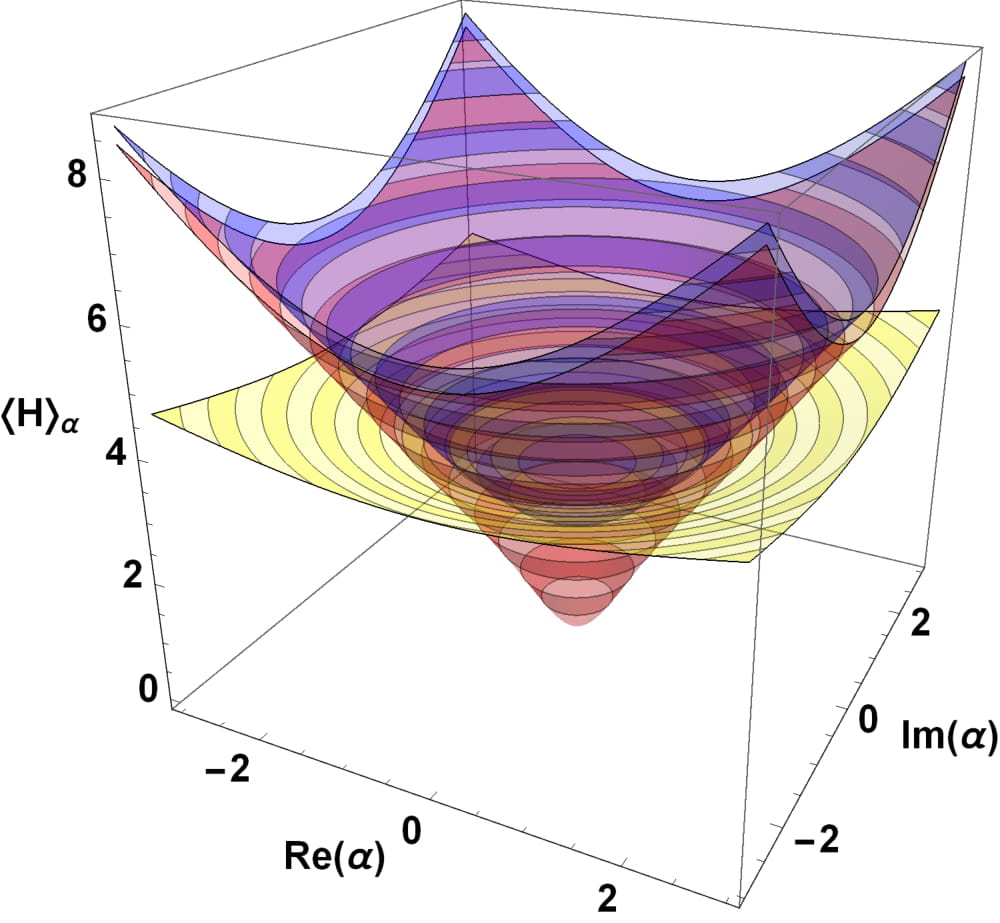}\\
		\centering{\footnotesize (b) $B_{0}=2$.}
		\label{fig::Halpha_strain_b}
	\end{minipage}
	\caption{\label{fig:Halpha_strain}Mean energy $\langle H\rangle_\alpha^\zeta/\hbar\sqrt{v_{xx}v_{yy}}$ as function of $\alpha$ for the nonlinear coherent states $\Psi_{\alpha}^f$: Eq.~(\ref{41}) (red), (\ref{46}) (blue) and (\ref{52}) (yellow). In all these cases, we take $\delta=0$.}
\end{figure}

\noindent while the mean energy $\langle H\rangle_\alpha^\zeta$ is (see Fig.~\ref{fig:Halpha_strain}):
\begin{equation}\label{46}
\langle H\rangle_\alpha^\zeta=\sqrt{v_{xx}v_{yy}}\langle H\rangle_\alpha, \quad \langle H\rangle_\alpha=\frac{\sqrt{\omega_{B}}\,\hbar}{\exp\left(\vert\tilde{\alpha}\vert^2\right)}\sum_{n=0}^{\infty}\frac{\vert\tilde{\alpha}\vert^{2n}}{n!}\sqrt{n+1},
\end{equation}
where $\langle H\rangle_\alpha$ is the corresponding mean energy for the pristine case for the same function $g(N)$.

	\begin{figure}[h!]
	\centering
	\begin{minipage}[b]{0.45\textwidth}
		\includegraphics[width=\textwidth]{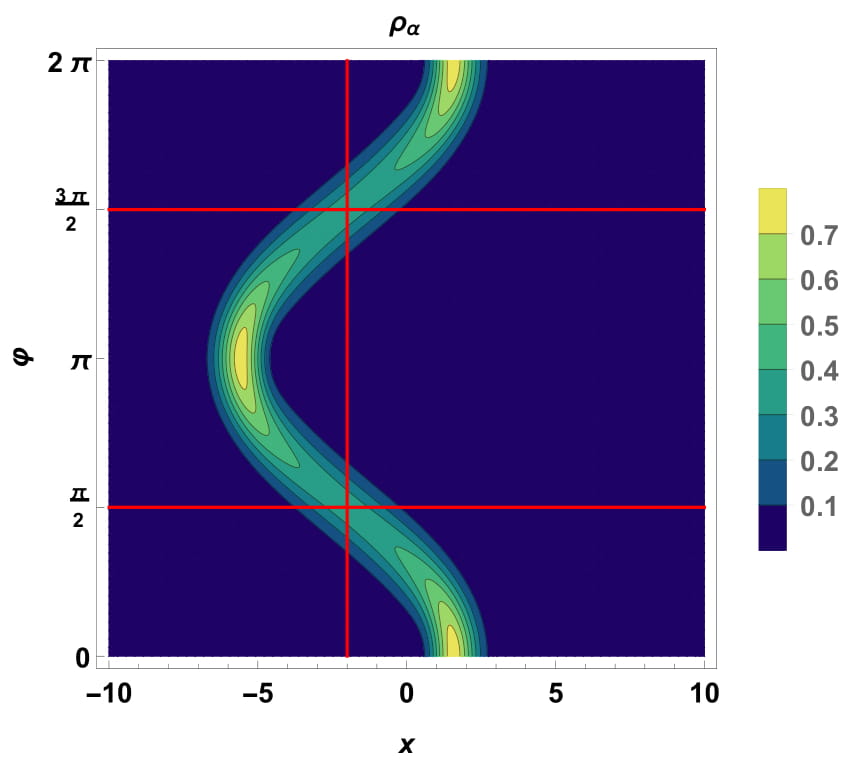}\\
		\centering{\footnotesize (a) $\zeta=1/2$.}
		\label{fig:rhoIIIA}
	\end{minipage}
	\hspace{1cm}
	~ 
	\begin{minipage}[b]{0.45\textwidth}
		\includegraphics[width=\textwidth]{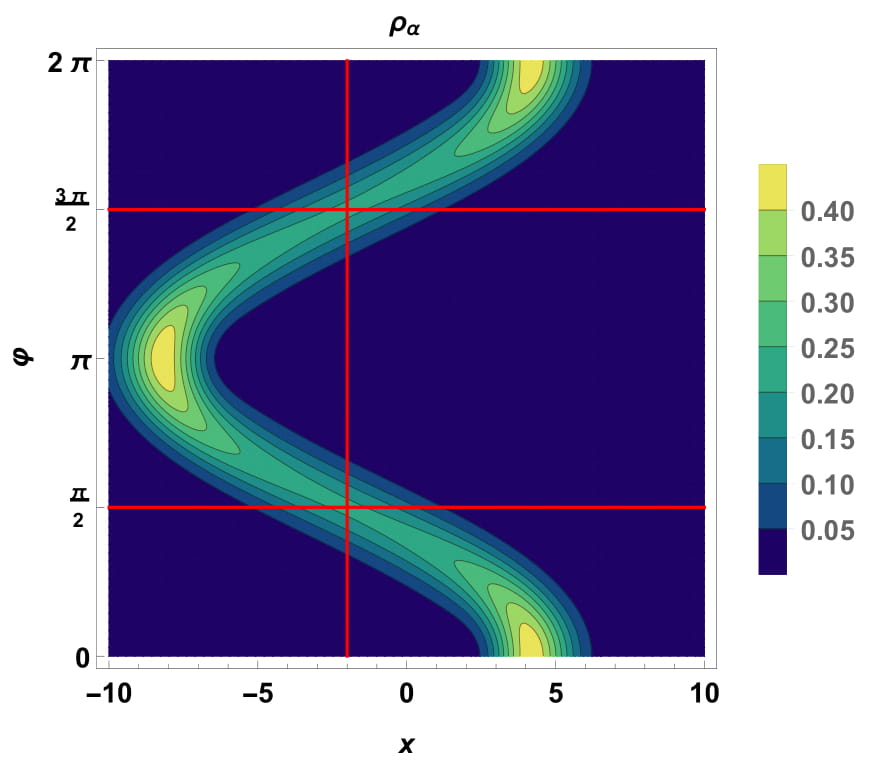}\\
		\centering{\footnotesize (b) $\zeta=3/2$.}
		\label{fig:rhoIIIB}
	\end{minipage}
	\caption{\label{fig:rhoIIIa}Probability density $\rho_{\alpha}(x)$ of the coherent states $\Psi^f_{\alpha}(x,y)$ in Eq.~(\ref{47}) for $\vert\alpha\vert=6$ and some values of the parameter $\zeta$. In all these cases, we take $B_0=1/2$, $k=\omega_{B}=1$ and $\delta=0$.}
\end{figure}

Analogously to the previous case, the center of the corresponding $\rho_{\alpha}(x)$ moves away from or approaches to the equilibrium position $x_0$ as $\vert\alpha\vert$ increases or decreases, respectively. By varying $\varphi\in[0,2\pi]$, the maximum probability performs again an oscillatory-like motion around $x_0$ (vertical red line in Fig.~\ref{fig:rhoIIa}), but when $\varphi=(2m+1)\pi/2$, $m=0,1,\dots$, $\rho_{\alpha}(x)$ is centered in such position (horizontal red lines in Fig.~\ref{fig:rhoIIa}). However, for small values of $\vert\alpha\vert$ and $\varphi=\pi/2$, the function $\rho_{\alpha}(x)$ decreases in the interjection of both lines, which is due to the behavior of the position dispersion $(\sigma_\xi)_{\alpha}$ for those values.

Moreover, the parameter $\zeta$ affects the probability density (see Fig.~\ref{fig:rhoII}): the value of $\rho_{\alpha}(x)$ increases when $v_{xx}$ decreases, while it tends to zero for $v_{xx}$ growing. Additionally, the center of the probability density is located either to the right, to the left or at the equilibrium point $x_0$ according to $0\leq\varphi<\pi/2$, $\pi/2<\varphi\leq2\pi$ or $\varphi=\pi/2$, respectively.

On the other hand, Fig.~\ref{fig:HURII} shows that the Heisenberg uncertainty relation reaches a maximum value equal to $1$ in the limit $\alpha\rightarrow0$, while it tends quickly to the minimum uncertainty value when $\alpha\rightarrow\infty$. In contrast to the previous case, this behavior is due to the state $\Psi_{1}(x,y)$, which is the minimum energy state that contributes to the corresponding superposition $\Psi_{\alpha}^f(x,y)$ in Eq.~(\ref{42}). Likewise, for values of $\vert\alpha\vert$ close to zero and $\varphi$ growing, the variances of the position $\xi$ and momentum $p$ operators change with respect to each other, becoming equal only when $\varphi=\pi/4$, but always being different to the usual value obtained for the standard coherent states of the harmonic oscillator. In particular, this implies that as $\vert\alpha\vert$ increases the uncertainty in the position reduces, as much as the quantum nature of such states allows.

\begin{figure}[h!]
	\centering
	\includegraphics[width=0.7\textwidth]{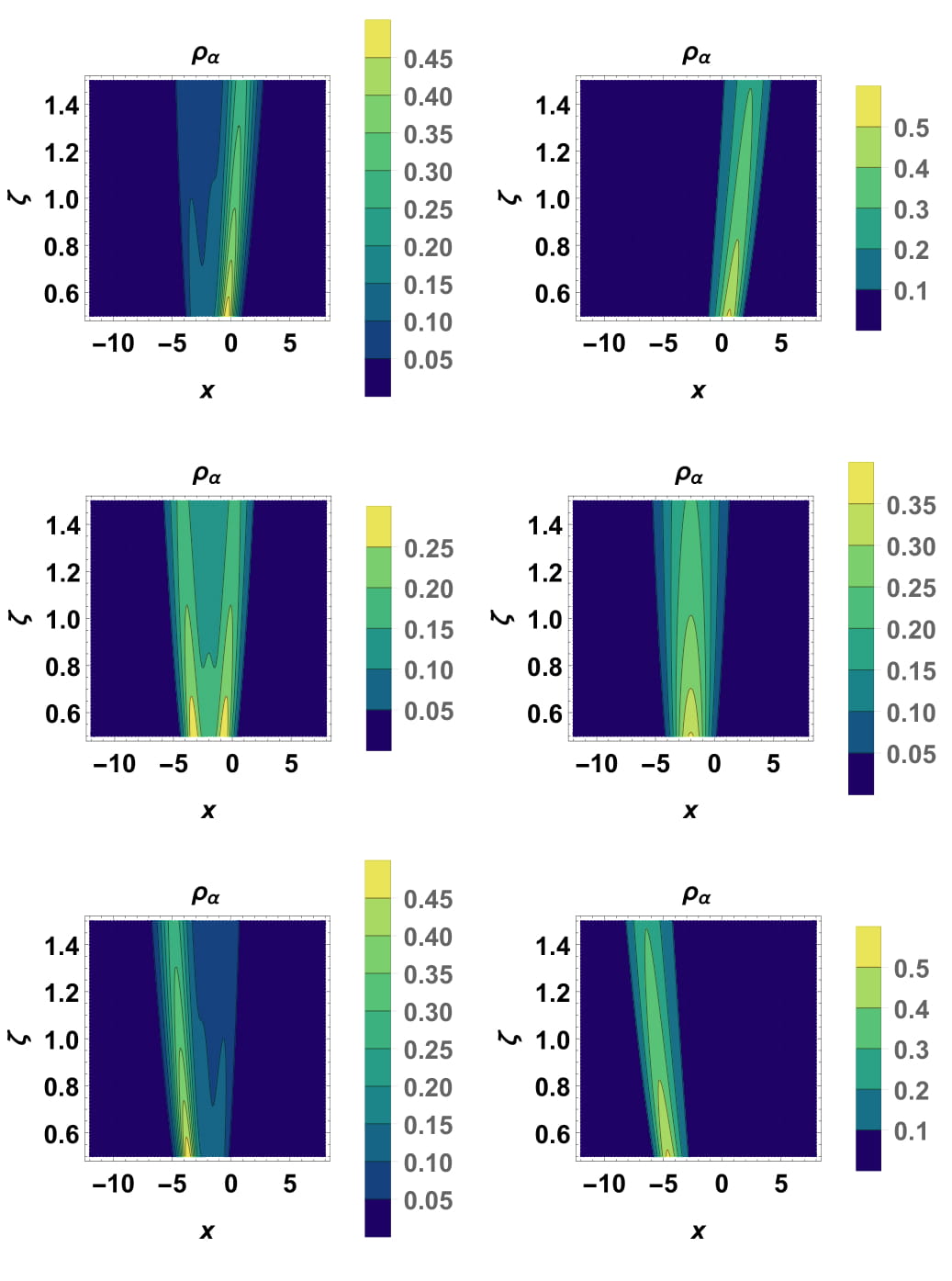}
	\caption{\label{fig:rhoIII}Probability density $\rho_{\alpha}(x)$ for the coherent states $\Psi^f_{\alpha}(x,y)$ in Eq.~(\ref{47}) as function of the parameter $\zeta$ for different values of eigenvalue $\alpha=\vert\alpha\vert\exp\left(i\varphi\right)$: (vertical) $\vert\alpha\vert=1,5$, and (horizontal) $\varphi=\pi/4,\pi/2,3\pi/4$. In all these cases, we take $B_0=1/2$, $k=\omega_{B}=1$ and $\delta=0$.}
\end{figure}


\begin{figure}[h!]
	\centering
	\begin{minipage}[b]{0.48\textwidth}
		\includegraphics[width=\textwidth]{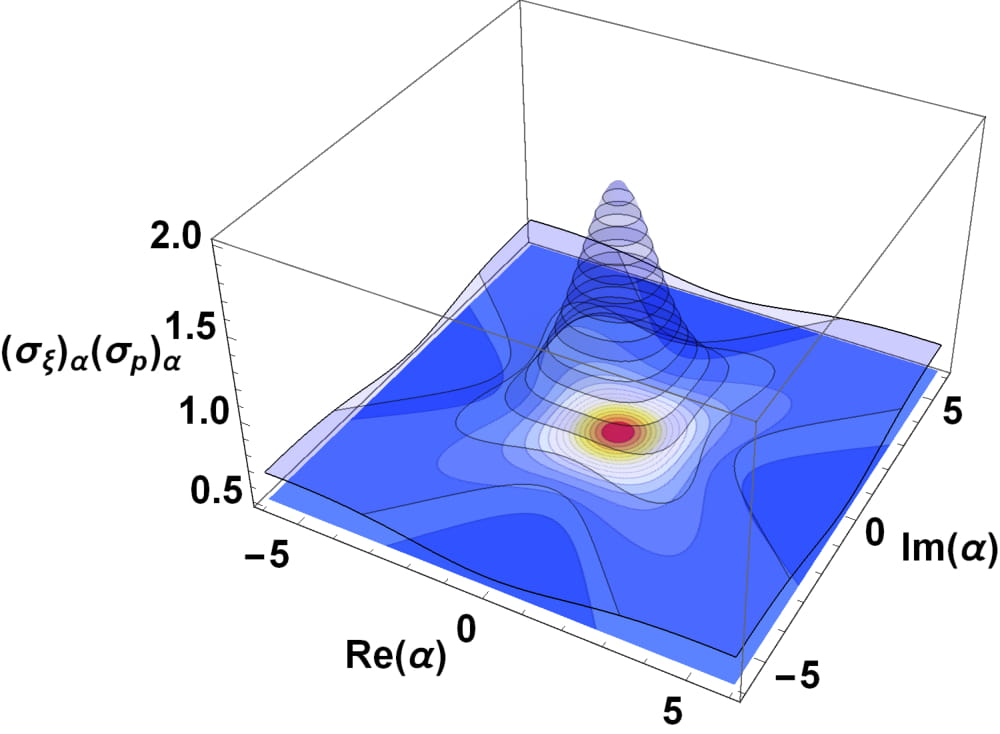}\\
		\centering{\footnotesize (a) $(\sigma_\xi)_\alpha(\sigma_p)_\alpha$.}
		\label{fig:HURIII1}
	\end{minipage}
	\hspace{0.7cm}
	~ 
	\begin{minipage}[b]{0.41\textwidth}
		\includegraphics[width=\textwidth]{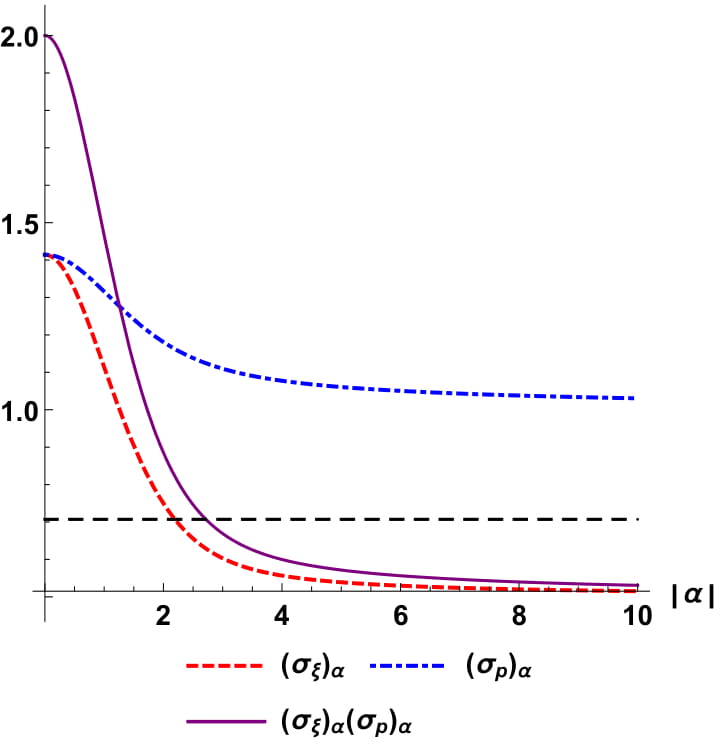}\\
		\centering{\footnotesize (b) $\varphi=0$.}
		\label{fig:HURIII2}
	\end{minipage}
	\hspace{1cm}
	~ 
	\begin{minipage}[b]{0.41\textwidth}
		\includegraphics[width=\textwidth]{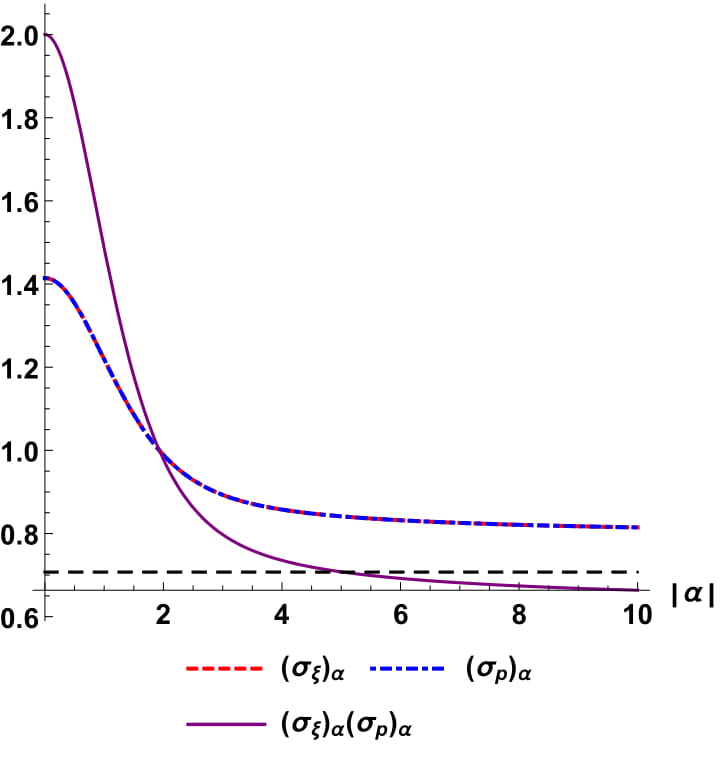}\\
		\centering{\footnotesize (c) $\varphi=\pi/4$.}
		\label{fig:HURIII3}
	\end{minipage}
	\hspace{1.5cm}
	~ 
	\begin{minipage}[b]{0.41\textwidth}
		\includegraphics[width=\textwidth]{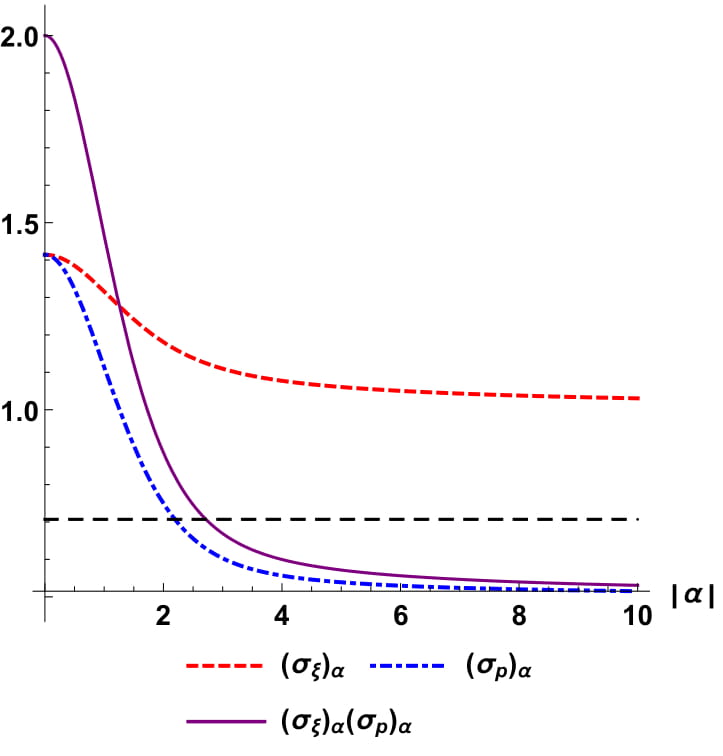}\\
		\centering{\footnotesize (d) $\varphi=\pi/2$.}
		\label{fig:HURIII4}
	\end{minipage}
	\caption{\label{fig:HURIII}For the states in Eq.~(\ref{47}): (a) $(\sigma_\xi)_\alpha(\sigma_p)_\alpha$ as function of $\alpha$.~(b-d) Comparison between $(\sigma_\xi)_\alpha$, $(\sigma_p)_\alpha$ and $(\sigma_\xi)_\alpha(\sigma_p)_\alpha$ as function of $\vert\alpha\vert$. Also, as  $\varphi$ changes, the dispersion of the position $\xi$ is lower, equal or upper than that of the momentum $p$.}
\end{figure}

\paragraph{b) $f(2)=0$}
Finally, for this case we consider the function $f(N+2)=h(N)=\sqrt{N}\times\sqrt{N+1}/\sqrt{N+2}$, which satisfies the condition $f(2)=0$. The corresponding NLCS are given by
\begin{equation}\label{47}
\Psi_{\alpha}^f(x,y)=\left(\frac{\vert\tilde{\alpha}\vert}{I_1(2\vert\tilde{\alpha}\vert)}\right)^{1/2}\sum_{n=0}^{\infty}\frac{\tilde{\alpha}^n}{\sqrt{n!(n+1)!}}\Psi_{n+2}(x,y),
\end{equation}
where $I_1(x)$ denotes the Bessel function of first kind. The probability density is then (see Figs.~\ref{fig:rhoIIIa} and \ref{fig:rhoIII})
\begin{eqnarray}\label{51}
\nonumber\rho_{\alpha}(x)&=&\Psi_{\alpha}^f(x,y)^\dagger\Psi_{\alpha}^f(x,y)=\left(\frac{\vert\tilde{\alpha}\vert}{2\,I_1(2\vert\tilde{\alpha}\vert)}\right)\left[\left\vert\sum\limits_{n=0}^{\infty}\frac{\tilde{\alpha}^n}{\sqrt{n!(n+1)!}}\psi_{n+2}(x)\right\vert^2\right.\nonumber\\
&&\left.+\left\vert\sum\limits_{n=0}^{\infty}\frac{\tilde{\alpha}^n}{\sqrt{n!(n+1)!}}\psi_{n+1}(x)\right\vert^2\right].
\end{eqnarray}

\begin{figure}[h!]
	\centering
	\begin{minipage}[b]{0.85\textwidth}
		\includegraphics[width=\textwidth]{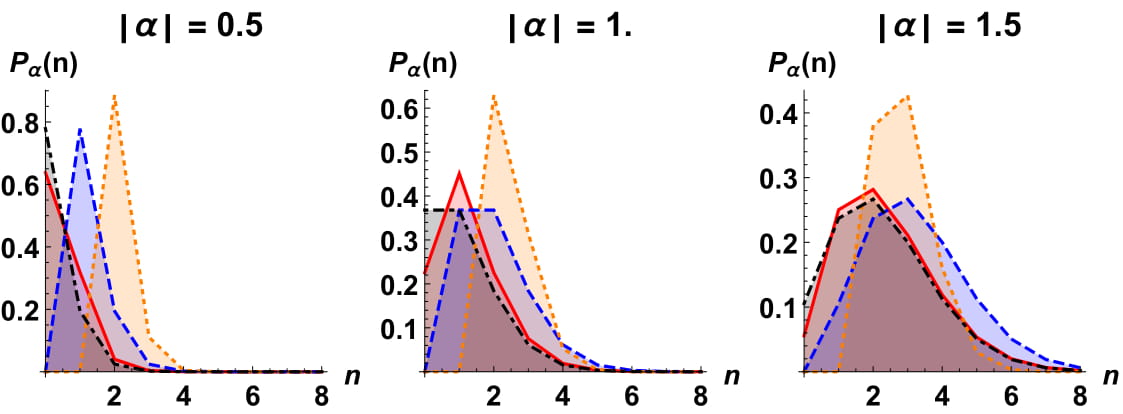}
		\label{fig:DCSA}
	\end{minipage}
	\hspace{0.7cm}
	~ 
	\begin{minipage}[b]{0.85\textwidth}
		\includegraphics[width=\textwidth]{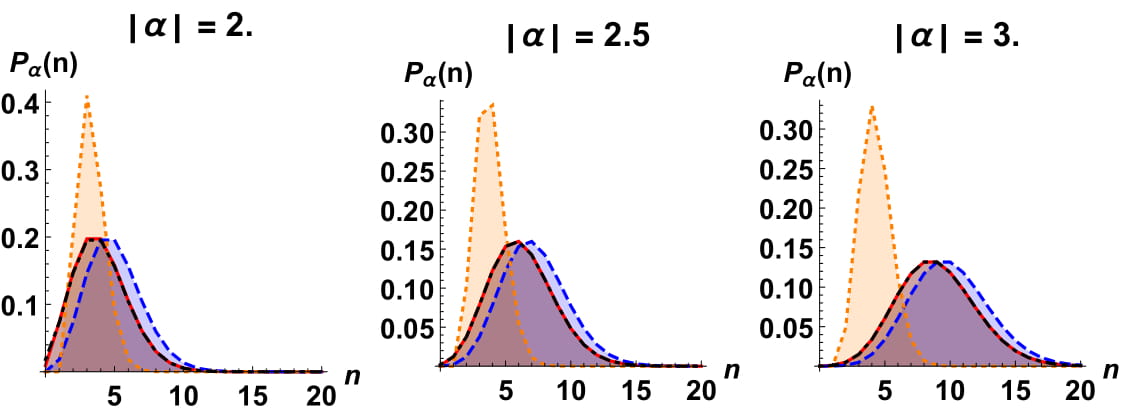}
		\label{fig:DCSB}
	\end{minipage}
	\caption{\label{fig:DCS}Occupation number distribution $P_{\alpha}(n)$ of each nonlinear coherent state $\Psi_{\alpha}^f$ --Eq.~(\ref{37}) (red, \full), (\ref{42}) (blue, \dashed), (\ref{47}) (orange, \dotted) and the Poisson distribution with $\lambda=\vert\alpha\vert^2$ (black, \chain)-- is shown for different values of $\vert\alpha\vert$. In all these cases, we take $\delta=0$.}
\end{figure}

Also, the quantities $\langle\mathbb{S}_q\rangle_{\alpha}$ and $\langle\mathbb{S}^2_q\rangle_{\alpha}$ are (see Fig.~\ref{fig:HURIII}):
\begin{subequations}
	\begin{align}
	\langle\mathbb{S}_q\rangle_{\alpha}&=\frac{\tilde{\alpha}+(-1)^q\tilde{\alpha}^\ast}{2\sqrt{2}i^q}\left(\frac{\vert\tilde{\alpha}\vert}{I_1(2\vert\tilde{\alpha}\vert)}\right)\left[\sum_{n=0}^{\infty}\frac{\vert\tilde{\alpha}\vert^{2n}}{\sqrt{n![(n+1)!]^3}}+\sum_{n=0}^{\infty}\frac{\sqrt{n+3}\,\vert\tilde{\alpha}\vert^{2n}}{\sqrt{n!(n+2)!}(n+1)!}\right], \\
	\nonumber\langle\mathbb{S}^2_q\rangle_{\alpha}&=2+\vert\tilde{\alpha}\vert\frac{I_2(2\vert\tilde{\alpha}\vert)}{I_1(2\vert\tilde{\alpha}\vert)}+(-1)^q\frac{(\tilde{\alpha}^2+\tilde{\alpha}^{\ast 2})}{4}\left(\frac{\vert\tilde{\alpha}\vert}{I_1(2\vert\tilde{\alpha}\vert)}\right)\times\\
	&\quad\times\left[\sum_{n=0}^{\infty}\frac{\vert\tilde{\alpha}\vert^{2n}}{\sqrt{n!(n+2)!}(n+1)!}+\sum_{n=0}^{\infty}\frac{\sqrt{n+4}\,\vert\tilde{\alpha}\vert^{2n}}{\sqrt{n!(n+1)!}(n+2)!}\right],
	\end{align}
\end{subequations}
and the mean energy $\langle H\rangle_\alpha^\zeta$ is given by (see Fig.~\ref{fig:Halpha_strain}):
\begin{equation}\label{52}
\langle H\rangle_\alpha^\zeta=\sqrt{v_{xx}v_{yy}}\langle H\rangle_\alpha, \quad \langle H\rangle_\alpha=\frac{\sqrt{\omega_{B}}\,\hbar\vert\tilde{\alpha}\vert}{I_1(2\vert\tilde{\alpha}\vert)}\sum_{n=0}^{\infty}\frac{\vert\tilde{\alpha}\vert^{2n}}{n!(n+1)!}\sqrt{n+2},
\end{equation}
where $\langle H\rangle_\alpha$ is the corresponding mean energy for the pristine case for the function $h(N)$.

Once again, the parameter $\zeta$ affects the probability density $\rho_{\alpha}(x)$ in Eq.~(\ref{51}) in a similar manner to the previous cases, changing also the center of such function with respect to the equilibrium position according to the value of $\varphi\in[0,2\pi]$ (see Fig.~\ref{fig:rhoIIIa}). However, while the position $x$ of the center of the probability density $\rho_{\alpha}(x)$ along the $x$-axis also changes with respect to $x_0$ due to the values of $\alpha=\vert\alpha\vert\exp(i\varphi)$ (vertical and horizontal red lines in Fig.~\ref{fig:rhoIIIa}), the distance between the points $x$ and $x_0$ is smaller in comparison with the cases already discussed (see Fig.~\ref{fig:rhoIII}). In other words, these states describe the particle motion close to $x_0$ even if $\zeta$ increases.

Furthermore, Fig.~\ref{fig:HURIII} shows that the behavior of the Heisenberg uncertainty relation associated to the states in Eq.~(\ref{47}) and variances of the position $\xi$ and momentum $p$ operators are different compared with the previous cases. Now, the HUR reaches a maximum value equal to $2$ in the limit $\alpha\rightarrow0$ but for $\alpha\rightarrow\infty$ it tends very slowly to $1/2$. This behavior is because the state $\Psi_{2}(x,y)$ is the minimum energy state that appears in the linear combination of $\Psi_{\alpha}^f(x,y)$ and so these NLCS cannot be considered as minimum uncertainty states. However, the behavior of the variances of both $\xi$ and $p$ operators in the limit $\vert\alpha\vert\rightarrow\infty$, suggest a squeezed-like behavior for them.

\begin{figure}[h!]
	\centering
	\begin{minipage}[b]{0.56\textwidth}
		\includegraphics[width=\textwidth]{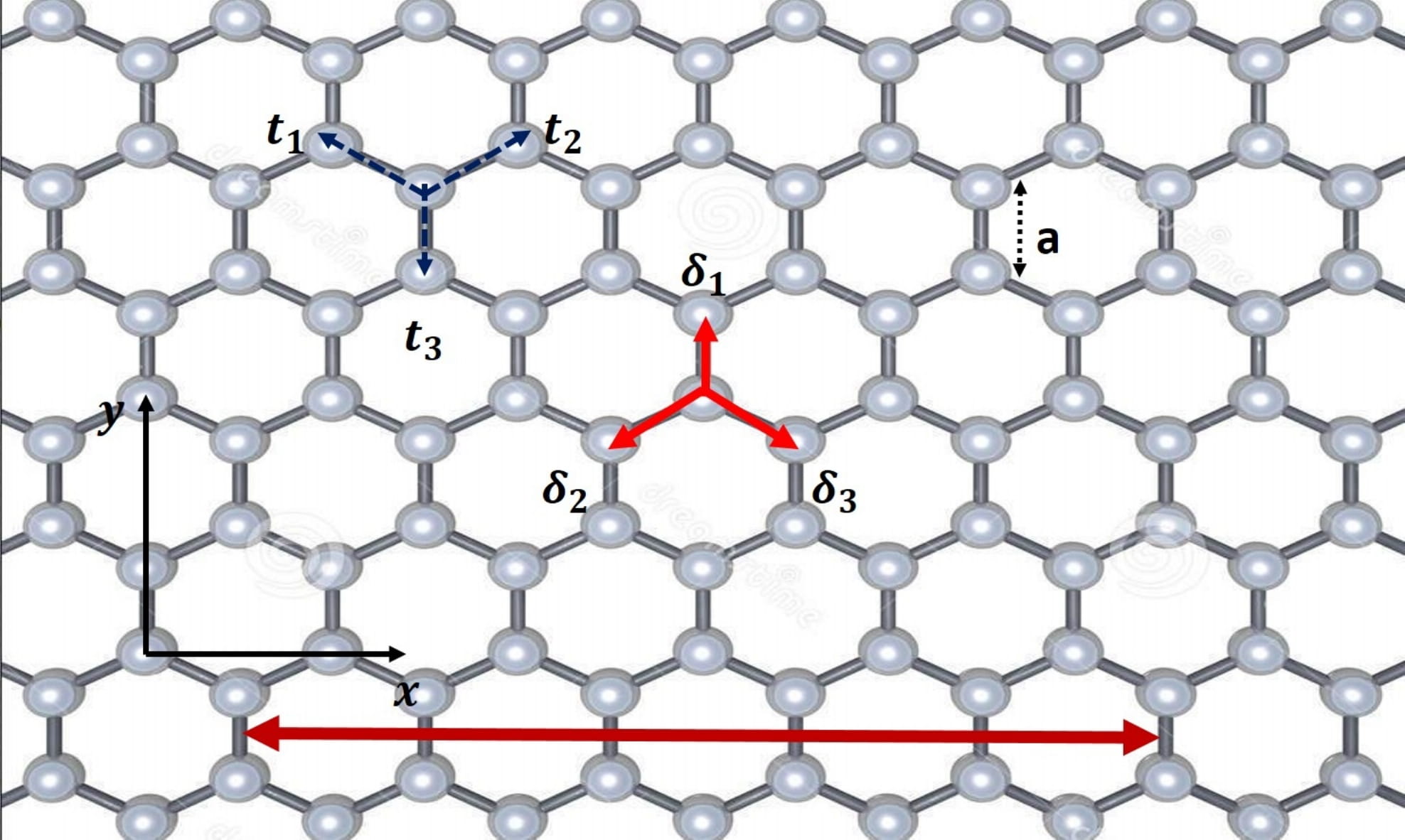}\\
		\centering{\footnotesize (a) $v_{xx}<v_{yy}$}
		\label{fig:latticea}
	\end{minipage}
	\hspace{0.1cm}
	\begin{minipage}[b]{0.37\textwidth}
		\includegraphics[width=\textwidth]{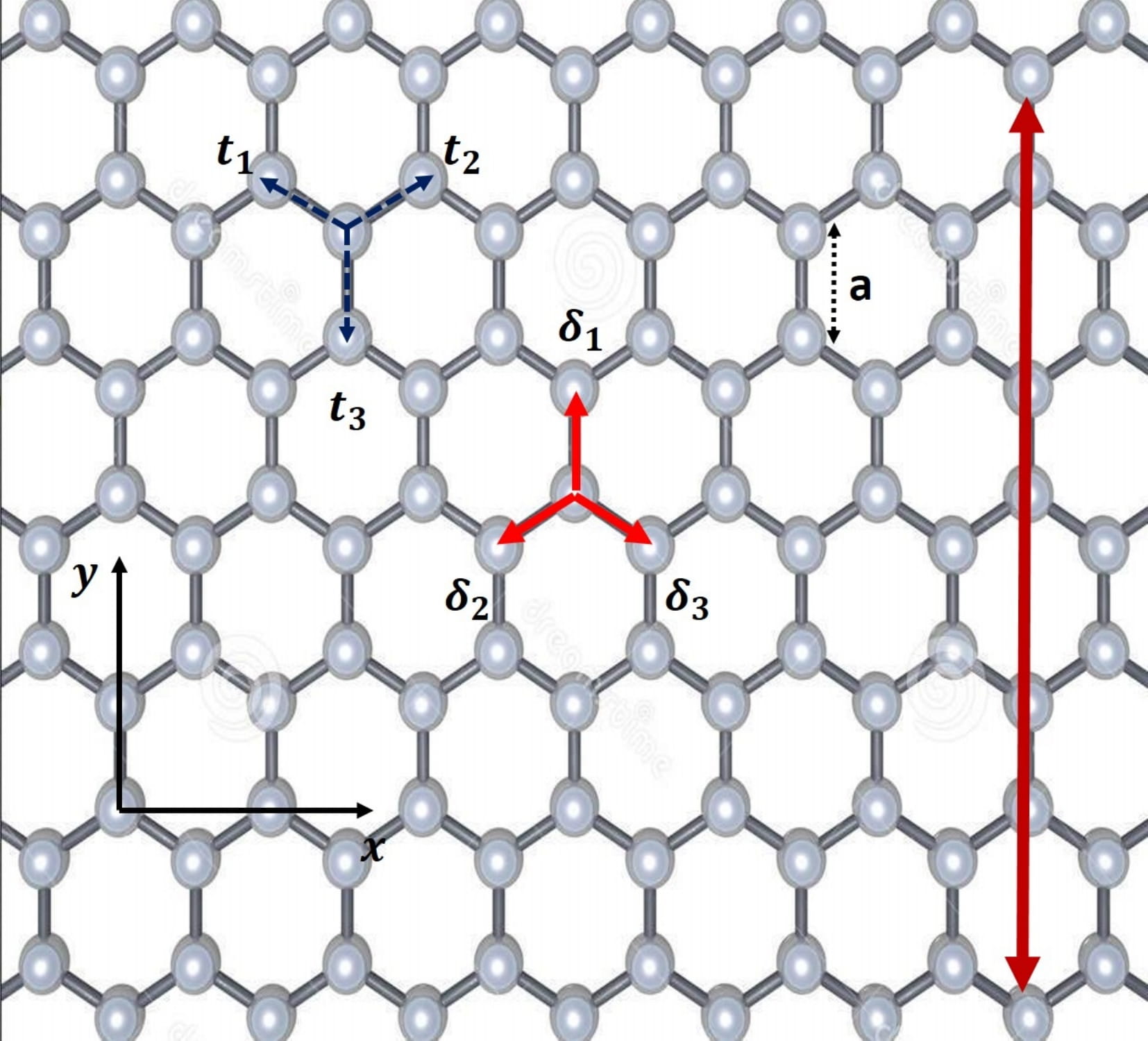}\\
		\centering{\footnotesize (b) $v_{xx}>v_{yy}$}
		\label{fig:latticec}
	\end{minipage}
	\caption{\label{fig:lattice}Honeycomb lattice for 2D-DM under uniform stress (red arrow) applied in a) the {\it zigzag} direction ($x$-axis, left panel) and b) the {\it armchair} direction ($y$-axis, right panel). Here, $\delta_{j}$ denotes the nearest neighbor vectors, $\mathsf{a}$ is the carbon-carbon distance and $t_j$ is hopping energy. For the pristine case in graphene, $\mathsf{a}_0\approx1.42$~{\AA} and $t_0\simeq2.7$ eV.}
\end{figure}

Finally, Fig.~\ref{fig:Halpha_strain} shows a comparison between the mean energy $\langle H\rangle_\alpha^\zeta$ corresponding to each NLCS $\Psi_{\alpha}^f(x,y)$ above described, while in Fig.~\ref{fig:DCS} the occupation number distribution $P_{\alpha}(n)=\vert\langle\Psi_{n}\vert\Psi_{\alpha}^f\rangle\vert^2\propto\vert a_{n}\vert^2$ of each NLCS is compared against the Poisson distribution with mean $\lambda=\vert\alpha\vert^2$, which is typical in the harmonic oscillator coherent states. As we can see, each mean energy is a continuous function of the eigenvalue $\alpha$ and in the limit, $\alpha\rightarrow0$ their behaviors are different due to the minimum energy state $\Psi_{n}(x)$ that contributes to the respective NLCS. Moreover, the mean energy is modified by the values of the velocities $v_{xx}$ and $v_{yy}$ due to the strain. On the other hand, the behavior of each distribution $P_{\alpha}(n)$ indicates that due to the form of the chosen function $f(N)$ for the states in Eq.~(\ref{47}), the probability distribution of its states does does not obey a Poisson-like distribution as $\vert\alpha\vert$ increases, in contrast with the other two cases for which the Poisson-like distribution is fulfilled.

\section{Discussion and conclusions}\label{conclu}

In this work, we have considered anisotropic 2D-Dirac-Weyl fermion systems immersed in a perpendicular uniform magnetic field, in order to explore the effects that the Dirac cones anisotropy has in the behavior of the nonlinear coherent states, which can be obtained by describing the background field in a  Landau-like gauge. This setup supplies a semi-classical description of the effects that the anisotropy have on the dynamics of the Dirac particles in a magnetic field. For our purposes, the anisotropy is characterized by the quantity $\zeta=v_{xx}/v_{yy}$, that indicates the anisotropy direction.

In what follows and for the sake of illustration, we consider as the 2D-DM a sample of strained graphene, in which for a uniform uniaxial strain~\cite{ow17,og13,bccc15,ow17-1,pap11}, the velocities $v_{ij}$ take the explicit form (see Fig.~\ref{fig:lattice}):
\small
\begin{subequations}
	\begin{itemize}
		\item if the uniaxial strain is applied along the $x$-direction:
		\begin{equation}
		v_{xx}=v_F(1-\beta\epsilon), \quad v_{yy}=v_F(1+\beta\nu\epsilon) \quad \Longrightarrow \quad \zeta\approx1-\beta(1+\nu)\epsilon+\mathcal{O}(\epsilon^2),
		\end{equation}
		\item if the uniaxial strain is applied along the $y$-direction:
		\begin{equation}
		v_{xx}=v_F(1+\beta\nu\epsilon), \quad v_{yy}=v_F(1-\beta\epsilon) \quad \Longrightarrow \quad \zeta\approx1+\beta(1+\nu)\epsilon+\mathcal{O}(\epsilon^2),
		\end{equation}
	\end{itemize}
\end{subequations}
\normalsize
where $v_F$ is the Fermi velocity of pristine graphene, $\beta\approx2$, $\epsilon$ indicates the strength of the applied strain  and $\nu$ is the Poisson ratio, which takes values in the range $\nu\sim0.1-0.15$. Hence, we can assume that the figures in the previous sections were obtained for graphene with $\nu=0.15$ and a strain of 21\% ($\epsilon=0.21$). On the other hand,
although the velocities $v_{ij}$ can be related with the strain tensor $\dvec{\epsilon}$ in a more general way, the uniform strain~\cite{og13,bccc15,ow17-1} deserves special attention, since it is the limiting case of any general deformation, is solvable and leads to an anisotropic Fermi velocity, but it does not produce any pseudo-magnetic field whatsoever. Due to its theoretical simplicity, we will use it for describing the effects induced in the dynamics of graphene electrons.

Thus, if $\zeta<1$, {\it i.e.}, $v_{xx}<v_{yy}$, the deformation takes place along the $x$-direction due to the interatomic distance $\mathsf{a}$ increases in the $x$-direction and the velocity $v_{xx}$ decreases, since the hopping energy $t$ also decreases (Fig.~\ref{fig:lattice}a). As a consequence of the Heisenberg uncertainty principle, the probability density of the NLCS is larger in comparison with the opposite case, $\zeta>1$, {\it i.e.}, $v_{xx}>v_{yy}$, in which the strain is applied along the $y$-direction because now the interatomic distance $\mathsf{a}$ decreases in the $x$-direction and the hopping energy $t$ increases (Fig.~\ref{fig:lattice}b). It means that when a uniform stress is applied on such a 2D-DM layer along the $x$-axis, we can think that the electrons are restricted to move in such direction and the probability to find them in a small interval in the $x$-axis increases because their velocity $v_{xx}$ decreases, while if the material is deformed in the orthogonal direction, the region where the electrons can be found increases and the probability decreases as $\zeta$, or $v_{yy}$, grows. In comparison with the pristine graphene case, where $v_{xx}=v_{yy}$, previous works~\cite{ed17,dnn19} show that the probability density can be modified by increasing or decreasing magnetic fields intensities but, due to the symmetry between the $x$ and $y$-coordinates, there is not a preferential direction for the restricted motion. However, as we can see in this work, by applying strain in either zigzag or armchair direction, one can talk about the {\em confinement} of the Dirac fermions in a particular direction because the material isotropic character is modified. In a sense, one could try to meet this fact with that shown in~\cite{og15}, where position-dependent Fermi velocities affect the probability densities.

In addition, from a semi-classical point of view, the eigenvalue $\alpha=\vert\alpha\vert\exp\left(i\varphi\right)$ somehow establishes an initial condition for the coherent states: for $\vert\alpha\vert$-values close to zero, the maximum probability  is found around the point $x_0$ and the effect of the strain is milder than when $\vert\alpha\vert$ is larger, allowing to localize the maximum probability away from the point $x_0$. In addition, if the center of  $\rho_{\alpha}(x)$ is located to the left ($\varphi>\pi/2$) or to the right ($\varphi<\pi/2 $) of the point $x_0$, when a deformation is applied along the $x$-axis, the distance between those points increases in the respective direction. It is important to remark that the CS obtained for the case $f(1)=f(2)=0$ tend to stay localized around the point $x_0$ even if the velocity $v_{xx}$ grows due to the strain applied in the armchair direction. Also, these states show a squeezed-like behavior because the ground state $\Psi_{0}$ is absent in the corresponding superposition and the form of the function $f(N+1)$ chosen. Recalling that different values of $\zeta=v_{xx}/v_{yy}$ change the shape of the quadratic potentials $V^{\pm}_{\zeta}(x)$ in (\ref{13}), it is clear that when $v_{xx}<v_{yy}$, the points of return $x$ approximate to $x_0$, so that the nonlinear coherent states in (\ref{47}) could better describe this situation (Fig.~\ref{fig:rhoIa}), while when $v_{xx}>v_{yy}$, the points $x$ move away from $x_0$ and the NLCS (\ref{37}) and (\ref{42}) could be used in this case (Figs.~\ref{fig:rhoIIa} and \ref{fig:rhoIIIa}), {\it i.e.}, as the amount $\zeta$ changes, we can choose from among these families of coherent states to better describe the problem according to the distance regime given by $\vert\alpha\vert$. Finally, the behavior of the occupation number distribution $P_{\alpha}(n)$ of each nonlinear coherent state compared with the Poisson distribution --that characterizes the eigenstates in the standard scalar coherent states-- allows us to conclude some important facts about them. For instance, the probability distribution of the states $\Psi_{n}$ in the first two families of NLCS obeys a Poisson-like distribution for growing $\vert\alpha\vert$-values, even in the coherent states $\Psi_{\alpha}^{f}$ in which there is no contribution of the Landau level $n=0$ (Eq.~(\ref{47})), in contrast with the third NLCS family for which $P_{\alpha}(n)$ does not fulfill a Poisson-like distribution. This fact is intimately related with the function $f(N)$ chosen in each case described and it is a sign of which coherent states would be easier to obtain experimentally, as occurs with the Gaussian wave packets which are characterized precisely by a Poisson distribution.  Nevertheless, we must not discard the idea that for some other forms of the function $f(N)$, perhaps the NLCS that do have or not the contribution of all the Landau level eigenstates could obey some other statistical distribution for any $\vert\alpha\vert$-value that also allows to easily obtain them in lab.

Since coherent states have been used in many branches of physics~\cite{ks85,zfg90,aag00,h14}, as in condensed matter physics~\cite{fk70} and atomic and molecular physics~\cite{acg72,wh73}, to analyze some measurable physical quantities, for experimental considerations we believe that the results obtained in this article can be useful to explore and describe phenomena on 2D-DM, perhaps of interference nature, because such a description establishes a bridge towards the phase space formalism that has been also employed in condensed matter physics~\cite{bbbj99,hhm16,ddms18}. Moreover, coherent states approach can be also extended to the description of crossed electric-magnetic fields effects, titled anisotropic Dirac cones and quantum electronics employing the Wigner function. Moreover, an alternative description of our finding can be obtained assuming a symmetric gauge for the background field, in order to describe either the bidimensional effects of the anisotropy on 2D materials lying on the $xy$-plane or by considering the problem where the velocities $v_{ij}$ can depend on the spatial coordinates. These studies are in progress and will be reported elsewhere. 

\section*{Acknowledgments}
The authors acknowledge David J. Fern\'andez and Maurice Oliva-Leyva for valuable discussions and careful reading of the manuscript. YCS acknowledges support from CIC-UMSNH under grant 3820801. AR acknowledges support from Consejo Nacional de Ciencia y Tecnolog\'{\i}a (M\'exico) under grant 256494. EDB acknowledges IFM-UMSNH for its warm hospitality and Act.~J. Manuel Zapata L. for giving the necessary impulse to continue researching.







\end{document}